\documentclass[twocolumn,letterpaper,aps,prd,longbibliography,superscriptaddress,showpacs,nofootinbib,floatfix]{revtex4-1}

\usepackage{graphicx}
\usepackage{amssymb}
\usepackage{amsmath}
\usepackage{xspace}

\newcommand{\gevc}{\mbox{GeV/$c$}\xspace}
\newcommand{\gevcsq}{\mbox{GeV/$c^{2}$}\xspace}

\newcommand{\raa}{\mbox{R$_{\rm CuAu}$}\xspace}
\newcommand{\pt}{\mbox{$p_{T}$}\xspace}
\newcommand{\jpsi}{\mbox{$J/\psi$}\xspace}
\newcommand{\jpsis}{\mbox{$J/\psi$s}\xspace}
\newcommand{\psip}{\mbox{$\psi^{\prime}$}\xspace}

\newcommand{\pp}{\mbox{$p$+$p$}\xspace}
\newcommand{\cc}{\mbox{$c\bar{c}$}\xspace}
\newcommand{\bb}{\mbox{$b\bar{b}$}\xspace}
\newcommand{\full}{\mbox{$\sqrt{s_{_{NN}}}=$ 200 GeV}\xspace}
\newcommand{\dcar}{\mbox{$\textrm{DCA}_{\rm R}$}\xspace}
\newcommand{\bfrac}{\mbox{$\textrm{F}_{B{\rightarrow}J/\psi}$}\xspace}
\newcommand{\btojpsi}{\mbox{$B{\rightarrow}J/\psi$}\xspace}

\def\func#1{\left(#1\right)}

\begin{document}

\title{$B$-meson production at forward and backward rapidity in $p$+$p$ 
and Cu+Au collisions at $\sqrt{s_{_{NN}}}$=200 GeV}

\newcommand{\abilene}{Abilene Christian University, Abilene, Texas 79699, USA}
\newcommand{\augie}{Department of Physics, Augustana University, Sioux Falls, South Dakota 57197, USA}
\newcommand{\banaras}{Department of Physics, Banaras Hindu University, Varanasi 221005, India}
\newcommand{\barc}{Bhabha Atomic Research Centre, Bombay 400 085, India}
\newcommand{\baruch}{Baruch College, City University of New York, New York, New York, 10010 USA}
\newcommand{\bnlcoll}{Collider-Accelerator Department, Brookhaven National Laboratory, Upton, New York 11973-5000, USA}
\newcommand{\bnlphys}{Physics Department, Brookhaven National Laboratory, Upton, New York 11973-5000, USA}
\newcommand{\caucr}{University of California-Riverside, Riverside, California 92521, USA}
\newcommand{\charlesczech}{Charles University, Ovocn\'{y} trh 5, Praha 1, 116 36, Prague, Czech Republic}
\newcommand{\chonbuk}{Chonbuk National University, Jeonju, 561-756, Korea}
\newcommand{\ciae}{Science and Technology on Nuclear Data Laboratory, China Institute of Atomic Energy, Beijing 102413, People's Republic of China}
\newcommand{\cns}{Center for Nuclear Study, Graduate School of Science, University of Tokyo, 7-3-1 Hongo, Bunkyo, Tokyo 113-0033, Japan}
\newcommand{\colorado}{University of Colorado, Boulder, Colorado 80309, USA}
\newcommand{\columbia}{Columbia University, New York, New York 10027 and Nevis Laboratories, Irvington, New York 10533, USA}
\newcommand{\czechtech}{Czech Technical University, Zikova 4, 166 36 Prague 6, Czech Republic}
\newcommand{\debrecen}{Debrecen University, H-4010 Debrecen, Egyetem t{\'e}r 1, Hungary}
\newcommand{\elte}{ELTE, E{\"o}tv{\"o}s Lor{\'a}nd University, H-1117 Budapest, P{\'a}zm{\'a}ny P.~s.~1/A, Hungary}
\newcommand{\eszterhazy}{Eszterh\'azy K\'aroly University, K\'aroly R\'obert Campus, H-3200 Gy\"ngy\"os, M\'atrai \'ut 36, Hungary}
\newcommand{\ewha}{Ewha Womans University, Seoul 120-750, Korea}
\newcommand{\fsu}{Florida State University, Tallahassee, Florida 32306, USA}
\newcommand{\gsu}{Georgia State University, Atlanta, Georgia 30303, USA}
\newcommand{\hanyang}{Hanyang University, Seoul 133-792, Korea}
\newcommand{\hiroshima}{Hiroshima University, Kagamiyama, Higashi-Hiroshima 739-8526, Japan}
\newcommand{\howard}{Department of Physics and Astronomy, Howard University, Washington, DC 20059, USA}
\newcommand{\ihepprot}{IHEP Protvino, State Research Center of Russian Federation, Institute for High Energy Physics, Protvino, 142281, Russia}
\newcommand{\illuiuc}{University of Illinois at Urbana-Champaign, Urbana, Illinois 61801, USA}
\newcommand{\inrras}{Institute for Nuclear Research of the Russian Academy of Sciences, prospekt 60-letiya Oktyabrya 7a, Moscow 117312, Russia}
\newcommand{\instpasczech}{Institute of Physics, Academy of Sciences of the Czech Republic, Na Slovance 2, 182 21 Prague 8, Czech Republic}
\newcommand{\isu}{Iowa State University, Ames, Iowa 50011, USA}
\newcommand{\jaea}{Advanced Science Research Center, Japan Atomic Energy Agency, 2-4 Shirakata Shirane, Tokai-mura, Naka-gun, Ibaraki-ken 319-1195, Japan}
\newcommand{\jyvaskyla}{Helsinki Institute of Physics and University of Jyv{\"a}skyl{\"a}, P.O.Box 35, FI-40014 Jyv{\"a}skyl{\"a}, Finland}
\newcommand{\kek}{KEK, High Energy Accelerator Research Organization, Tsukuba, Ibaraki 305-0801, Japan}
\newcommand{\korea}{Korea University, Seoul, 136-701, Korea}
\newcommand{\kurchatov}{National Research Center ``Kurchatov Institute", Moscow, 123098 Russia}
\newcommand{\kyoto}{Kyoto University, Kyoto 606-8502, Japan}
\newcommand{\labllr}{Laboratoire Leprince-Ringuet, Ecole Polytechnique, CNRS-IN2P3, Route de Saclay, F-91128, Palaiseau, France}
\newcommand{\lahorelums}{Physics Department, Lahore University of Management Sciences, Lahore 54792, Pakistan}
\newcommand{\lawllnl}{Lawrence Livermore National Laboratory, Livermore, California 94550, USA}
\newcommand{\losalamos}{Los Alamos National Laboratory, Los Alamos, New Mexico 87545, USA}
\newcommand{\lund}{Department of Physics, Lund University, Box 118, SE-221 00 Lund, Sweden}
\newcommand{\lyon}{IPNL, CNRS/IN2P3, Univ Lyon, Universit{\'e} Lyon 1, F-69622, Villeurbanne, France}
\newcommand{\maryland}{University of Maryland, College Park, Maryland 20742, USA}
\newcommand{\mass}{Department of Physics, University of Massachusetts, Amherst, Massachusetts 01003-9337, USA}
\newcommand{\michigan}{Department of Physics, University of Michigan, Ann Arbor, Michigan 48109-1040, USA}
\newcommand{\muhlenberg}{Muhlenberg College, Allentown, Pennsylvania 18104-5586, USA}
\newcommand{\myongji}{Myongji University, Yongin, Kyonggido 449-728, Korea}
\newcommand{\nagasaki}{Nagasaki Institute of Applied Science, Nagasaki-shi, Nagasaki 851-0193, Japan}
\newcommand{\nara}{Nara Women's University, Kita-uoya Nishi-machi Nara 630-8506, Japan}
\newcommand{\natmephi}{National Research Nuclear University, MEPhI, Moscow Engineering Physics Institute, Moscow, 115409, Russia}
\newcommand{\newmex}{University of New Mexico, Albuquerque, New Mexico 87131, USA}
\newcommand{\nmsu}{New Mexico State University, Las Cruces, New Mexico 88003, USA}
\newcommand{\ohio}{Department of Physics and Astronomy, Ohio University, Athens, Ohio 45701, USA}
\newcommand{\ornl}{Oak Ridge National Laboratory, Oak Ridge, Tennessee 37831, USA}
\newcommand{\orsay}{IPN-Orsay, Univ.~Paris-Sud, CNRS/IN2P3, Universit\'e Paris-Saclay, BP1, F-91406, Orsay, France}
\newcommand{\peking}{Peking University, Beijing 100871, People's Republic of China}
\newcommand{\pnpi}{PNPI, Petersburg Nuclear Physics Institute, Gatchina, Leningrad region, 188300, Russia}
\newcommand{\riken}{RIKEN Nishina Center for Accelerator-Based Science, Wako, Saitama 351-0198, Japan}
\newcommand{\rikjrbrc}{RIKEN BNL Research Center, Brookhaven National Laboratory, Upton, New York 11973-5000, USA}
\newcommand{\rikkyo}{Physics Department, Rikkyo University, 3-34-1 Nishi-Ikebukuro, Toshima, Tokyo 171-8501, Japan}
\newcommand{\saispbstu}{Saint Petersburg State Polytechnic University, St.~Petersburg, 195251 Russia}
\newcommand{\seoulnat}{Department of Physics and Astronomy, Seoul National University, Seoul 151-742, Korea}
\newcommand{\stonybrkc}{Chemistry Department, Stony Brook University, SUNY, Stony Brook, New York 11794-3400, USA}
\newcommand{\stonycrkp}{Department of Physics and Astronomy, Stony Brook University, SUNY, Stony Brook, New York 11794-3800, USA}
\newcommand{\sungskku}{Sungkyunkwan University, Suwon, 440-746, Korea}
\newcommand{\tenn}{University of Tennessee, Knoxville, Tennessee 37996, USA}
\newcommand{\titech}{Department of Physics, Tokyo Institute of Technology, Oh-okayama, Meguro, Tokyo 152-8551, Japan}
\newcommand{\tsukuba}{Center for Integrated Research in Fundamental Science and Engineering, University of Tsukuba, Tsukuba, Ibaraki 305, Japan}
\newcommand{\vandy}{Vanderbilt University, Nashville, Tennessee 37235, USA}
\newcommand{\weizmann}{Weizmann Institute, Rehovot 76100, Israel}
\newcommand{\wigner}{Institute for Particle and Nuclear Physics, Wigner Research Centre for Physics, Hungarian Academy of Sciences (Wigner RCP, RMKI) H-1525 Budapest 114, POBox 49, Budapest, Hungary}
\newcommand{\yonsei}{Yonsei University, IPAP, Seoul 120-749, Korea}
\newcommand{\zagreb}{Department of Physics, Faculty of Science, University of Zagreb, Bijeni\v{c}ka c.~32 HR-10002 Zagreb, Croatia}
\affiliation{\abilene}
\affiliation{\augie}
\affiliation{\banaras}
\affiliation{\barc}
\affiliation{\baruch}
\affiliation{\bnlcoll}
\affiliation{\bnlphys}
\affiliation{\caucr}
\affiliation{\charlesczech}
\affiliation{\chonbuk}
\affiliation{\ciae}
\affiliation{\cns}
\affiliation{\colorado}
\affiliation{\columbia}
\affiliation{\czechtech}
\affiliation{\debrecen}
\affiliation{\elte}
\affiliation{\eszterhazy}
\affiliation{\ewha}
\affiliation{\fsu}
\affiliation{\gsu}
\affiliation{\hanyang}
\affiliation{\hiroshima}
\affiliation{\howard}
\affiliation{\ihepprot}
\affiliation{\illuiuc}
\affiliation{\inrras}
\affiliation{\instpasczech}
\affiliation{\isu}
\affiliation{\jaea}
\affiliation{\jyvaskyla}
\affiliation{\kek}
\affiliation{\korea}
\affiliation{\kurchatov}
\affiliation{\kyoto}
\affiliation{\labllr}
\affiliation{\lahorelums}
\affiliation{\lawllnl}
\affiliation{\losalamos}
\affiliation{\lund}
\affiliation{\lyon}
\affiliation{\maryland}
\affiliation{\mass}
\affiliation{\michigan}
\affiliation{\muhlenberg}
\affiliation{\myongji}
\affiliation{\nagasaki}
\affiliation{\nara}
\affiliation{\natmephi}
\affiliation{\newmex}
\affiliation{\nmsu}
\affiliation{\ohio}
\affiliation{\ornl}
\affiliation{\orsay}
\affiliation{\peking}
\affiliation{\pnpi}
\affiliation{\riken}
\affiliation{\rikjrbrc}
\affiliation{\rikkyo}
\affiliation{\saispbstu}
\affiliation{\seoulnat}
\affiliation{\stonybrkc}
\affiliation{\stonycrkp}
\affiliation{\sungskku}
\affiliation{\tenn}
\affiliation{\titech}
\affiliation{\tsukuba}
\affiliation{\vandy}
\affiliation{\weizmann}
\affiliation{\wigner}
\affiliation{\yonsei}
\affiliation{\zagreb}
\author{C.~Aidala} \affiliation{\losalamos} \affiliation{\michigan} 
\author{N.N.~Ajitanand} \altaffiliation{Deceased} \affiliation{\stonybrkc} 
\author{Y.~Akiba} \email[PHENIX Spokesperson: ]{akiba@rcf.rhic.bnl.gov} \affiliation{\riken} \affiliation{\rikjrbrc} 
\author{R.~Akimoto} \affiliation{\cns} 
\author{J.~Alexander} \affiliation{\stonybrkc} 
\author{M.~Alfred} \affiliation{\howard} 
\author{V.~Andrieux} \affiliation{\michigan} 
\author{K.~Aoki} \affiliation{\kek} \affiliation{\riken} 
\author{N.~Apadula} \affiliation{\isu} \affiliation{\stonycrkp} 
\author{H.~Asano} \affiliation{\kyoto} \affiliation{\riken} 
% ...
\collaboration{PHENIX Collaboration} \noaffiliation


\begin{abstract}


The fraction of $J/\psi$ mesons which come from $B$-meson decay, 
$\textrm{F}_{B{\rightarrow}J/\psi}$, is measured for J/$\psi$ rapidity 
\mbox{$1.2<|y|<2.2$} and $p_T>0$ in $p$+$p$ and Cu+Au collisions at 
$\sqrt{s_{_{NN}}}$=200 GeV with the PHENIX detector. The extracted fraction 
is $\textrm{F}_{B{\rightarrow}J/\psi}$ = 0.025 $\pm$ 0.006(stat) $\pm$ 
0.010(syst) for $p$+$p$ collisions. For Cu+Au collisions, 
$\textrm{F}_{B{\rightarrow}J/\psi}$ is 0.094 $\pm$ 0.028(stat) $\pm$ 
0.037(syst) in the Au-going direction ($-2.2<y<-1.2$) and 0.089 $\pm$ 
0.026(stat) $\pm$ 0.040(syst) in the Cu-going direction ($1.2<y<2.2$). 
The nuclear modification factor, $R_{\rm CuAu}$, of $B$~mesons in Cu+Au 
collisions is consistent with binary scaling of measured yields in 
$p$+$p$ at both forward and backward rapidity.

\end{abstract}


\date{\today}


\maketitle

\section{Introduction.}
\label{sec:introduction}

Heavy quarks (charm and bottom quarks in the context of this work)
are a powerful tool to investigate initial-state effects 
and quark-gluon plasma (QGP) medium formation in heavy ion collisions. 
Initial-state nuclear effects can alter the number of heavy quarks 
produced compared to extrapolations from \pp collisions. Once produced, 
however, the number of heavy quarks is preserved in strong interactions 
in the medium.  Final-state effects, such as energy loss in the QGP 
\cite{Bjorken:1982tu,Dokshitzer2001199}, can only modify the momentum 
distribution of these quarks and open heavy mesons.  On the other hand, 
the number of prompt \jpsi is not expected to be preserved in the medium 
because of the low $c\bar{c}$ binding energy, which can allow the \jpsi 
to be broken by medium interactions.

The $B$~mesons which decay to \jpsi and subsequently decay into dimuon 
pairs, represent a relatively clean channel to extract $b$-quark yields.  
At the Relativistic Heavy Ion Collider (RHIC), the PHENIX forward 
silicon-vertex detector (FVTX) along with the central silicon-vertex 
detector (VTX) provide the ability to measure precise event vertex 
positions as well as displacement of the decay muon trajectories from 
the reconstructed event vertex.  This allows for the statistical 
separation of \jpsis from $B$-meson decays from prompt \jpsis. The 
PHENIX muon detector acceptance for $B$~mesons in this channel is nearly 
constant over all transverse momenta, allowing for the direct extraction 
of momentum integrated $b$-quark yields, corresponding to the number of 
$b$ quarks in the rapidity acceptance. With these yields, we can verify 
whether initial-state effects are relevant to $B$-meson production in 
nucleus+nucleus collisions.

The only known source of nonprompt \jpsi is production via $B$-meson 
decay \cite{Agashe:2014kda}, which has a typical decay time $\tau_0 
\sim$1.5--1.6 ps \cite{Aaij:2014owa}. The constituent bottom quarks have 
mass $m_b\sim$4.5 \gevcsq and are created from processes such as gluon 
fusion ($gg\rightarrow Q\bar{Q}$), flavor excitation where heavy quarks 
($Q$) from the nucleon wave function scatter with gluons ($Qg\rightarrow 
Qg$), and gluon splitting ($gg\rightarrow Q\bar{Q}g$) 
\cite{Norrbin2000,PhysRevD.74.054010}. Gluon-gluon fusion and flavor 
excitation are equally dominant whereas the gluon splitting contribution 
is small at $\sqrt{s}=$200 GeV according to {\sc pythia{\footnotesize 8}} 
hard scattering simulations \cite{Sjostrand:2007gs}.
	
Initial-state effects on the precursor gluons, before their hard 
scattering, include coherent multiple scattering, also called dynamical 
shadowing \cite{PhysRevD.74.054010}, incoherent multiple scattering 
\cite{Kang:2013ufa,Kang:2014hha}, initial state energy loss 
\cite{Kang:2015mta}, and saturation of small momentum fraction $x$ 
gluons \cite{McLerran:1993ka}. Nuclear parton distribution functions, 
extracted from deep inelastic scattering and Drell-Yan experimental 
data, such as EPS09 \cite{Eskola:2009uj} and impact-parameter-dependent 
EPS09s \cite{EPS09s}, indicate a pattern of suppression for small-$x$ 
gluons and an enhancement of intermediate to large-$x$ gluons. 
Semi-leptonic decays of particles carrying heavy flavor and produced
inclusively in $d$$+$Au 
collisions at \full \cite{Adare:2013lkk,Adare:2012yxa} indicate yield 
suppression at forward rapidity, where Bjorken $x\sim5\times 10^{-3}$ 
gluons dominate. The same analysis also revealed a clear yield 
enhancement at mid- and backward rapidity, where gluons with fractional 
momentum $0.05\lesssim x \lesssim 0.2$ are dominant. A similar 
enhancement was observed in the measurements of $D$ mesons in $d$$+$Au at 
midrapidity \cite{Adams:2004fc}.

Asymmetric Cu+Au collisions have the advantage of being a relatively 
large system which can access different $x$ regions at positive and 
negative rapidities. Studies in these collisions provide a powerful test 
of how initial-state effects observed in small systems, such as $p$$+A$, 
can be projected in large heavy ion collisions. In the PHENIX muon arms, 
positive rapidity corresponds to the Cu-going direction, probing 
small-$x$ in the Au nucleus and large-$x$ in the Cu. Negative rapidity 
covers the Au-going direction, probing small-$x$ in the Cu nucleus and 
large-$x$ in the Au. Initial state parton distribution modifications are 
predicted to be stronger in the Au nucleus \cite{Aidala:2014bqx}. If 
these modifications have the same pattern as seen in $p(d)$+A 
collisions, they may cause suppression in the Cu-going (positive 
rapidity) direction and enhancement in the Au-going direction. 
Initial-state energy loss \cite{Xing201277} also results in larger 
suppression at positive rapidity.

In this study we quantify the fraction of muons from the decay of 
nonprompt \jpsi in \pp and Cu+Au collisions at \full using the PHENIX 
muon arms measuring in the rapidity range $1.2<|y|<2.2$. The nonprompt 
fractions are then used to calculate the nuclear modification of 
$B$~mesons in Cu+Au collisions. This result uses the capability to 
measure the approach between muons and the collision vertex using the 
FVTX~\cite{Aidala201444}. Section~\ref{sec:apparatus_data_set} describes 
the experimental apparatus and the data set. 
Section~\ref{sec:data_selection} describes the data selection, 
backgrounds and defines the distance of closest approach (\dcar). 
Section~\ref{sec:MC_setup} describes the simulation setup used to obtain 
the \dcar distribution profiles. The fit to the real data distance of 
closest approach distributions is detailed in Section 
\ref{sec:fitting_procedure}. Systematic uncertainties are discussed in 
Section~\ref{sec:sys_errors}. Results and interpretations are presented 
in Section~\ref{sec:results}.

\section{Experimental Apparatus and Data Set.}
\label{sec:apparatus_data_set}

\begin{figure}[t]
	\includegraphics[width=1.0\linewidth]{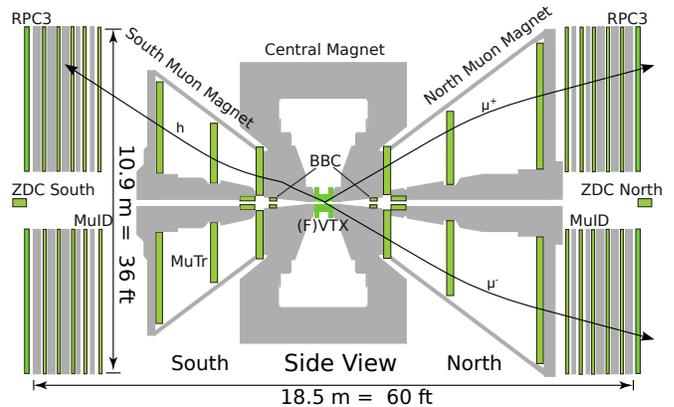}
\caption{\label{fig:phenix_2012}Schematic view of the detector apparatus 
along with an illustrated dimuon and a hadron stopping in one of the 
muon identifier (MuID) gaps.}
\end{figure}

This analysis was performed using data sets obtained with the PHENIX 
detector (Fig.~\ref{fig:phenix_2012}) at the Relativistic Heavy Ion Collider 
during the 2015 \pp and 
2012 Cu+Au \full runs.  Each collision was identified by the north and 
south beam-beam counters (BBC) which each comprise 64 quartz radiators 
instrumented with mesh dynode PMTs covering charged particles in the 
pseudorapidity region $3.0<|\eta|<3.9$ and with time resolution of 
$\sim50$ ps. A fast online collision $z$ vertex can be determined from 
the difference between the average north and south arrival times. A 
minimum-bias (MB) event is triggered by two or more hits in each BBC and 
a measured vertex position $|z_{\rm vtx}|\lesssim10$ cm, and results in 
an acceptance of $93\pm3\%$ of the total Cu+Au cross section. For \pp 
collisions, one or more hits are required in each BBC for the MB 
trigger, and $55\pm5\%$ of the total inelastic cross section is 
accepted. This analysis also used a sample of dimuon triggered events 
(MUIDLL1\_2D) which required two roads of hits found in the Iarocci 
tubes in the muon identifiers (MuID), including at least one road that 
reached the most downstream tubes.

\subsection{PHENIX muon spectrometers.}
\label{sec:muon_arms}

Each muon spectrometer covers a pseudorapidity range of $1.2<|\eta|<2.2$ 
and $2\pi$ in azimuth. Each spectrometer comprises hadron-absorber 
material, muon trackers (MuTr) inside a conical-shaped magnet, and a 
MuID.  The first layer of hadron absorber, placed 
between the FVTX and the MuTr, comprises 19 cm of copper, 60 cm of 
iron from the central magnet and 36.2 cm steel, corresponding to a total 
of 7.2 nuclear interaction lengths. This material absorbs pions and 
kaons emitted into the acceptance of the muon arms. The muon magnet 
system provides a radial field inside the MuTr volume of approximately 
$\int \mathbf{B\cdot dl}=0.72$ T$\cdot$m at 15 degrees from the beam 
axis, bending particles in the azimuthal direction.

Tracking and momentum measurements are performed by the MuTr, which 
comprises 8 octants of cathode strip chambers distributed in each of 
three $z$-stations. The first two stations have three sensitive planes 
each, and the farthest station from the interaction region has two 
sensitive planes. Each plane contains two cathode readout strip planes 
with varying stereo angle orientations among the planes in the station 
in order to provide measurements in two spatial dimensions. The momentum 
resolution achieved by the MuTr is $\delta p/p \approx 0.05$ for a 
typical muon from \jpsi decays.

The north and south MuID systems are located downstream of the MuTr and 
comprise five absorber plates totaling 4.8 (5.4) nuclear 
interaction lengths in the south (north) arm. Two Iarocci tube planes 
with vertical and horizontal orientations distributed in six individual 
panels are placed after each of the five absorber gaps. Pion and kaon 
rejection after all absorber material is larger than a factor of 250. 
Only muons with momentum $>3~\gevc$ are able to penetrate all absorbers. 
Recorded hits in the tubes are used to reconstruct roads which are used 
in the MUIDLL1\_2D trigger and in full muon track reconstruction. 
Technical details of the muon arms can be found in 
\cite{Akikawa2003537,Adler:2006yu}.

\label{sec:FVTX}

\begin{figure}[!ht]
\centering
\includegraphics[width=1.0\linewidth]{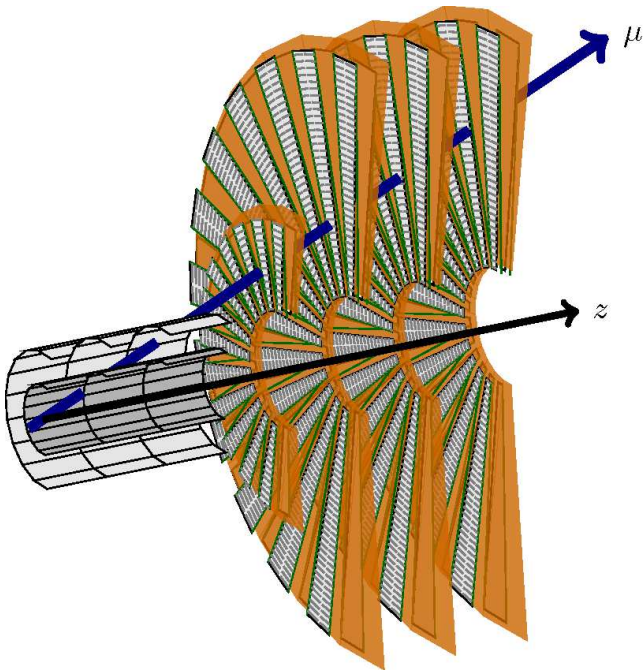}
\caption{Schematic view of the west VTX and north-west arm FVTX 
detectors used in this analysis.}
	\label{fig:fvtx}
\end{figure}

Particles produced at the primary vertex cross $\sim$7.2 interaction 
lengths of absorber before reaching the first MuTr station located at 
$z=\pm190$ cm. Due to multiple scattering in the absorber, the 
projection of tracks reconstructed in the MuTr to the FVTX has a 
standard deviation radius of 3~cm. The FVTX helps track these particles 
from the absorber to the vertex point.

A precise vertex measurement is provided offline 
(Sec.~\ref{sec:primary_vertex}) by the central arm vertex detector (VTX) 
and the FVTX. The VTX \cite{Li2004300,Ichimiya:2008am} is a silicon 
detector with four radial layers placed at 2.6, 5.1, 11.8 and 16.7 cm 
from the $z$-axis, covering $2\times\Delta\phi\approx$ 1.6$\pi$ and 
$|z_{\rm vtx}|<10$ cm. The innermost two layers have pixel segmentation 
of $\Delta\phi \times \Delta z=$ 50$~\mu$m$\times$425$~\mu$m and the two 
outer layers comprise stripixels with an effective pixel size of 80 
$\mu$m in R$\Delta\phi$ and 1 mm in z.

\begin{table}[ht]
\caption{\label{tab:FVTX_specs} Summary of FVTX geometry.}
\begin{ruledtabular}    \begin{tabular}{rl}
    Mean $z$-position of discs [cm] & \ $\pm$20.1, $\pm$26.1, $\pm$32.2, $\pm$38.2 \\
    $\phi$ segmentation each disk & \ 48 wedges $\times$ 2 columns \\
    inner radius & \ 4.4 cm\\
    strip pitch & \ 75 $\mu$m \\
	strip length & \ 3.4 mm to 11.5 mm \\
	total number of strips & \ 1.08 M \\
    silicon thickness & \ 320 $\mu$m \\
    \end{tabular}  \end{ruledtabular}
\end{table}

The FVTX is a silicon detector installed in 2012 to precisely measure 
the radial distance of closest approach (\dcar) of extrapolated particle 
trajectories to the collision vertex (Sec.~\ref{sec:DCA}), allowing 
statistical separation between prompt muons and muons from the decay of 
long-lifetime particles. Geometrical characteristics of the FVTX are 
listed in Table~\ref{tab:FVTX_specs}. More technical details concerning 
the FVTX detector can be found in \cite{Aidala201444}.

Tracks reconstructed in the silicon system are required to have at least 
three hits in different FVTX disks and/or VTX pixel layers (seen as 
half-cylinders in Fig.~\ref{fig:fvtx}). The magnetic field in the region 
of the FVTX is primarily in the $z$-direction, resulting in only a very 
small bending of tracks in the $\phi$ direction. Therefore, the FVTX 
cannot measure particle momentum, which can be reconstructed only if the FVTX 
track is associated to a MuTr track.

\subsection{Data set and quality assurance.}
\label{sec:data_set_qa}

Only collisions with a vertex determined by the VTX and FVTX within 
$z=\pm10$ cm from the nominal interaction point are selected. 
Collisions where the vertex determination 
(Sec.~\ref{sec:primary_vertex}) has an uncertainty larger than 500 
$\mu$m (200 $\mu$m) are also rejected in \pp (Cu+Au) data. The fraction 
of collisions within the vertex range used in this analysis is 14\% in 
\pp and 18\% in Cu+Au. Only MB triggers or MUIDLL1\_2D triggers in 
coincidence with a MB trigger (MUIDLL1\_2D\&MB) were analyzed. In \pp 
the number of analyzed events in the selected vertex region is 
$3.4\times 10^{11}$, corresponding to an integrated luminosity of 
$\int \mathcal{L} dt = 14.8~{\rm pb}^{-1}$. In Cu+Au 5.7$\times 10^{9}$ 
MB and 284$\times 10^{6}$ MUIDLL1\_2D\&MB events were analyzed in the 
selected vertex range. The MB trigger is sensitive to 
$\sigma_{Cu+Au}=$5.23$\pm$0.15 b of the total Cu+Au cross section, 
according to Glauber calculations reported in \cite{Aidala:2014bqx}. 
Therefore, the Cu+Au events used in this analysis correspond to $\int 
\mathcal{L} dt$ = 1.0 nb$^{-1}$, or the \pp equivalent of 11.8 
pb$^{-1}$.

The run-by-run average \dcar (described in Sec.\ref{sec:DCA}) measured 
for of all charged particles by the FVTX was found to be stable 
throughout the data collection period in the Cu+Au run. For the case of 
\pp collisions, $\sim10\%$ of data are rejected due to instabilities of 
the measured \dcar.

\section{Data Selection and How to Obtain the Distance of Closest 
Approach Distributions.}
\label{sec:data_selection}

B~mesons have a mean decay length $c\tau_0\sim450~\mu$m, a scale 
measurable by precision vertex detectors. At large rapidity the momentum 
boost $\gamma=\sqrt{p_z^2+p_T^2+m^2}/m$ to the decay length is larger 
than at midrapidity ($p_z\sim 0$), allowing $B$-meson identification via 
nonprompt decays even at zero transverse momentum.

Typically, the fraction of $B$-meson decays in \jpsi samples (here defined 
as \bfrac) is determined by measuring the vertex given by the intersection
of the two muon trajectories in the dimuon pair. This approach is not used 
in this analysis because: 1) the \jpsi dimuon vertex cannot be precisely 
determined because of the limited $\phi$ resolution of the FVTX; 2) a 
fraction of the sample of \jpsi decays would be lost in the sample when one
of the muons reconstructed by the MuTr does not match to the FVTX.

In this analysis the \jpsi decay is identified by the dimuon invariant 
mass of selected MuTr+MuID tracks. In a second step, muon candidates from the 
identified \jpsi decay that have matching FVTX tracks are selected. The 
radial distance of closest approach \dcar (Sec.~\ref{sec:DCA}) is 
determined for combined FVTX+MuTr+MuID tracks after quality cuts. The 
following sections detail how the sample is selected, how the 
backgrounds are treated, and how \dcar is defined.

\subsection{Selection of \jpsi candidates.}
\label{sec:jpsi_selection}

\begin{figure*}[ht]
    \includegraphics[width=0.99\linewidth]{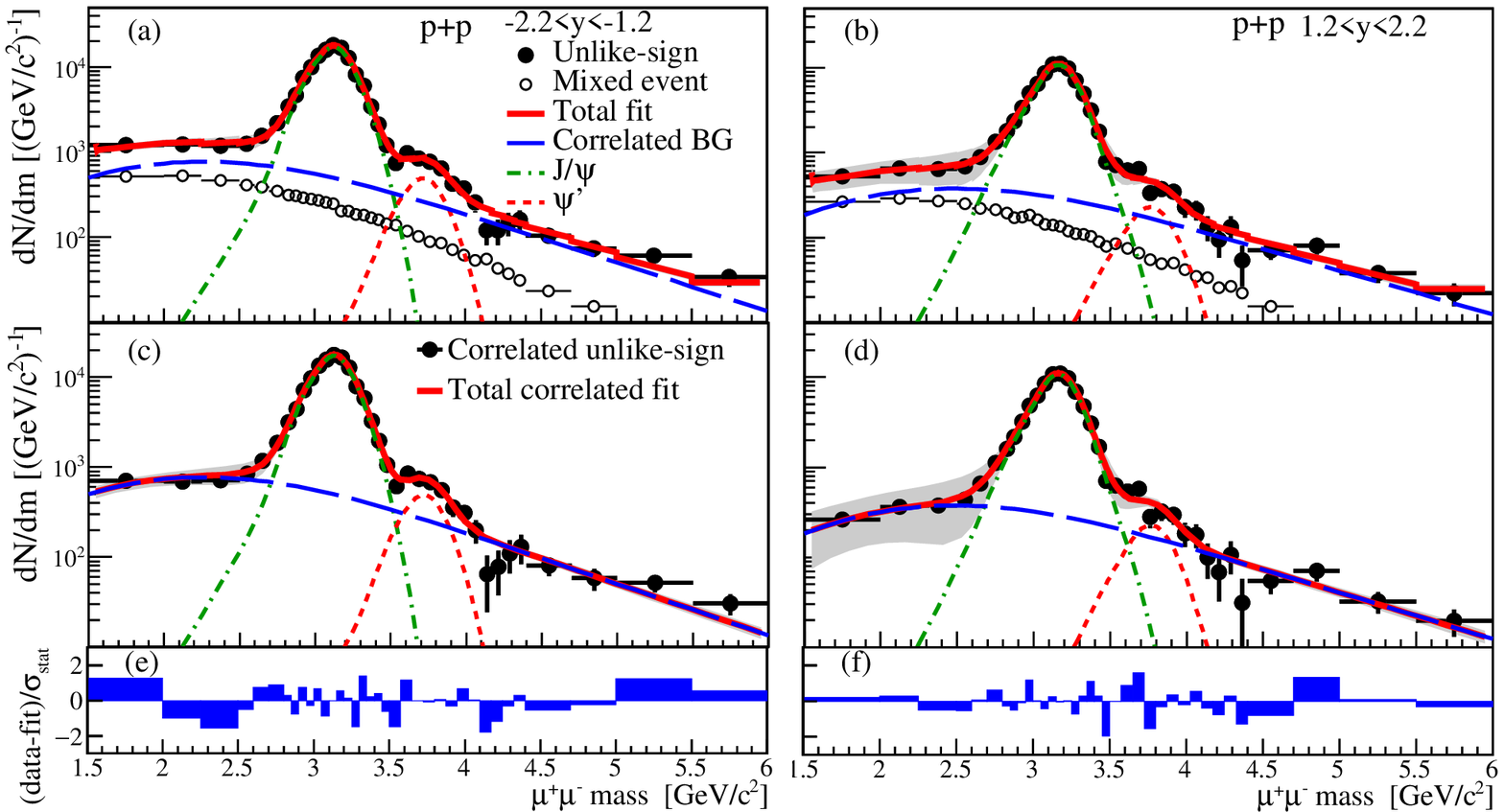}
    \includegraphics[width=0.99\linewidth]{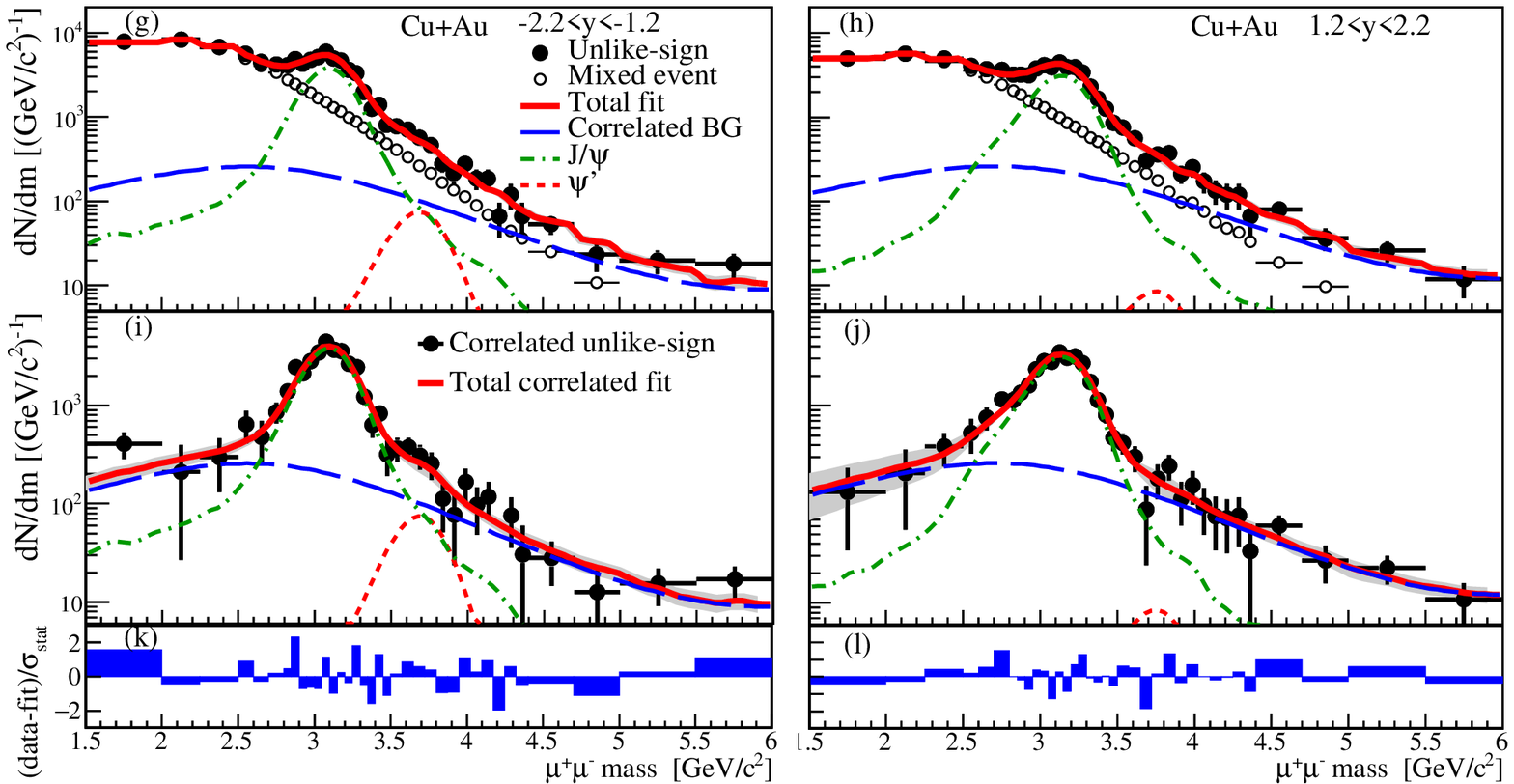}
\caption{The dimuon mass distributions in \pp (a-f) and Cu+Au (g-l) 
collisions. The mass distribution along with mixed event dimuons which 
account for the combinatorial background is shown in panels (a,b,g,h). 
The distribution after removal of the combinatorial background is shown 
in panels (c,d,i,j). The lines represent the 
Eq.~(\ref{eq:dimuon_mass_fit}) components fit to the data points. Panels 
(e,f,k,l) show the pull between data and fit.}
	\label{fig:mass_plots_pp}
\end{figure*}

Single muon candidates are MuTr tracks that are associated to MuID roads 
(Sec.~\ref{sec:muon_arms}). The MuTr track is required to have at least 
11 out of a maximum of 16 hits in different cathode planes, and the 
fitted track should return a $\chi^2/$NDF$<$10, cut based on the 
distribution obtained from simulated muons which also accounts for 
tracks crossing malfunctioning channels in the MuTr. Because a track is 
required to cross all absorbers, only muons with a total reconstructed 
momentum larger than 3 \gevc are accepted. The MuID roads are required 
to have at least 6 out of a possible 10 hits in different tube planes 
including a hit in one of the planes located behind the last MuID 
absorber. Three standard deviation cuts are applied to the distance and 
angle between the MuTr track and the MuID road projections at the first 
MuID gap. Dimuon pairs formed from MuTr+MuID selected muons are required 
to have an opening angle larger than 45$^o$ if the dimuon \pt is smaller 
than 5 \gevc. Implementation of this cut helps remove contributions from 
ghost tracks. A fit involving the muon pair tracks and the collision 
vertex is required to have a $\chi^2/{\rm NDF}<5(3)$ in \pp (Cu+Au) 
data, corresponding to a maximum distance between the dimuon crossing 
and the collision point of approximately 1~cm. This cut has no impact on 
$B$ decay acceptance.

Dimuon spectra in the region of the \jpsi mass are shown in Fig.\ 
\ref{fig:mass_plots_pp} for both muon arms. The combinatorial 
background, shown as open circles in panels (a) and (b), corresponds to 
random combinations of muons. This background is determined from mixed 
event dimuons and normalized by the geometric average between same event 
like-sign sources $N_{++}^{\rm same}$, $N_{--}^{\rm same}$ and mixed 
event like-sign sources $N_{++}^{\rm mix}$, $N_{--}^{\rm mix}$:

\begin{equation}
\label{eq:norm_cb}
\textrm{Norm}_{\rm CB} = \frac{\sqrt{N_{++}^{\rm same} \cdot N_{--}^{\rm same}}}{\sqrt{N_{++}^{\rm mix} \cdot N_{--}^{\rm mix}}}.
\end{equation}

Mixed event dimuons are required to come from two separate events with 
a $z$-vertex difference no larger than 1.5 cm and a difference between 
collision centrality percentiles that is no larger than 5\%. Accumulated 
mixed event dimuon counts are five times larger than the same-event 
dimuons in order to reduce background statistical uncertainties. The 
number of selected \jpsi decays inside the dimuon invariant mass $2.8< 
m_{\mu\mu} [\gevcsq] < 3.5$, the signal over background, after all 
quality cuts, and the fraction of correlated continuum background 
$f_{\rm cont}$ are listed in Table~\ref{tab:jpsi_counts}.

\begin{table}[!ht]
\caption{\label{tab:jpsi_counts}Dimuon net counts, signal to combinatorial 
background (S/CBG) and the correlated background contribution 
$f_{cont}$ in the \jpsi mass region $2.8< m_{\mu\mu} [\gevcsq] < 
3.5$.}
\begin{ruledtabular} 	\begin{tabular}{lcccc}
		data & arm & net count & S/CBG & $f_{\rm cont}$ \\ \hline
		\pp & south & 5978$\pm$150  & 32.7  &  5.5 $\pm$ 0.3\% \\
		\pp & north & 3714$\pm$67 & 33.6 &  5.3 $\pm$ 0.4\% \\ 
		Cu+Au & south & 3075$\pm$92 & 1.13 & 10.3 $\pm$ 2.4 \% \\
		Cu+Au & north & 2675$\pm$82 & 1.35 & 10.7 $\pm$ 2.5 \% \\
	\end{tabular}  \end{ruledtabular} 
\end{table}

Correlated background expected to contribute in the \jpsi mass region is 
mainly from \mbox{$c\bar{c} + X \rightarrow \mu^+\mu^-+X$} 
and \mbox{$b\bar{b} + X \rightarrow \mu^+\mu^-+X$} processes. 
The amount of correlated background shown as dashed lines in 
Fig.~\ref{fig:mass_plots_pp} is estimated by fitting the function

\begin{eqnarray}
	\label{eq:dimuon_mass_fit}
    f_{\mu^+\mu^-}(m) &=& {\rm CBG}(m) + N_{\rm corr }f_{\rm corr}(m) \\\nonumber
    &+& N_{J/\psi}f_{J/\psi}(m) + N_{\psi^{\prime}}f_{\psi^{\prime}}(m)\\\nonumber
    f_{\rm corr}(m) &=& A\varepsilon(m)\left[c_{\rm lm} e^{-m/\lambda_{lm}}+ 
\left(1-c_{\rm lm}\right)m^{-\lambda_{hm}}\right]
\end{eqnarray}

\noindent to the unlike-sign dimuon mass distribution $m$, where 
CBG$(m)$ is the combinatorial background, $f_{\rm corr}(m)$ is the 
correlated background, $f_{J/\psi}(m)$ and $f_{\psi^{\prime}}$ are the 
\jpsi and \psip simulated peaks, $A\varepsilon(m)$ is the mass dependent 
detector acceptance and efficiency determined from simulation and 
trigger emulator. The correlated background functional form of $f_{\rm 
corr}(m)$ accounts for an exponential behavior at low mass and a power 
law behavior at high mass, verified in {\sc pythia{\footnotesize 8}} 
simulation of \cc and \bb pair production. The free parameters in the 
fit are $N_{\rm corr}$, $N_{J/\psi}$, $N_{\psi^{\prime}}$, $c_{\rm lm}$, 
$\lambda_{\rm lm}$ and $\lambda_{\rm hm}$. Fitting constrains are 
applied to $c_{\rm lm}$, $\lambda_{\rm lm}$ and $\lambda_{\rm hm}$ 
according to a $f_{\rm corr}(m)$ fit to correlated like-sign dimuon mass 
distribution in the same mass range.  Table~\ref{tab:jpsi_counts} also 
shows the extracted fractions of correlated background in the \jpsi mass 
region:

\begin{equation}
	f_{\rm cont} = \int_{m=2.8}^{3.5} dm~N_{\rm corr}f_{\rm corr}(m).
\end{equation}

\subsection{FVTX-MuTr track association.}
\label{sec:FVTX_MuTr_association}

Requirements for standalone FVTX track selection include a minimum of 
three hits in different FVTX or VTX planes. The analysis require a FVTX 
track has a $\chi^2$ probability (p-value) larger than 5\%. The track 
quality selection keeps 95\% of the true tracks according to 
simulations. FVTX-MuTr track matching is performed as follows: 
\begin{enumerate}

\item MuTr and FVTX tracks are projected to three $z$-plane locations: 
1) the fourth FVTX disk from the vertex, 2) the middle of the absorber 
materials in front of the MuTr and 3) the first MuTr station. Radial 
($R$ and $p_{R}$) and azimuthal ($\phi$ and $p_{\phi}$) position and 
momentum projections, and uncertainties ($\sigma_{R_{\rm FVTX-MuTr}}$, 
$\sigma_{\phi_{\rm FVTX-MuTr}}$, $\sigma_{p_{R_{\rm FVTX-MuTr}}}$, 
$\sigma_{p_{\phi_{\rm FVTX-MuTr}}}$) are calculated for each FVTX and MuTr 
track using the GEANE Kalman Filter algorithm \cite{Innocente:247710} 
taking into account the detector geometry, the amount of absorber 
material and the magnetic field map. The momentum magnitude used for 
FVTX track projections is taken from the associated MuTr track.

\item A combined FVTX+MuTr track $\chi^2$ is calculated for each 
association from individual $\chi^2_{\rm FVTX}$ and $\chi^2_{\rm MuTr}$ 
track qualities and FVTX-MuTr projection residuals in the tree planes in 
an approximation considering the correlation between residuals in 
different planes is small:

\begin{equation}
    	\chi^2 =  \chi^2_{\rm FVTX} + \chi^2_{\rm MuTr} + \chi^2_{\rm match}
	\end{equation}
\begin{eqnarray*}
        \chi^2_{\rm match} 
= \sum_{\rm i=1}^{3} & \frac{\left(R_i^{\rm FVTX} 
- R_i^{\rm MuTr}\right)^2}{\sigma R^2_{\rm FVTX-MuTr}} +
\frac{\left(\phi_i^{\rm FVTX} 
- \phi_i^{\rm MuTr}\right)^2}{\sigma \phi^2_{\rm FVTX-MuTr}} +\\\nonumber
& \frac{\left(pr_i^{\rm FVTX} 
- pr_i^{\rm MuTr}\right)^2}{\sigma pr^2_{\rm FVTX-MuTr}} 
+ \frac{\left(p\phi_i^{\rm FVTX} 
- p\phi_i^{\rm MuTr}\right)^2}{\sigma p\phi^2_{\rm FVTX-MuTr}},\\\nonumber            
    \end{eqnarray*}

\noindent where $p\phi_i^{\rm FVTX}$ and $pr_i^{\rm FVTX}$ are just 
track directions determined by the FVTX.

\item Only FVTX-MuTr combined tracks with $\chi^2$/NDF$<$6(3) are 
accepted for further analysis in \pp(Cu+Au) data.

\item In the \pp analysis, at least one VTX hit is required for tracks 
inside the VTX acceptance to reduce effects from additional material 
that was installed just before the 2015 run between the VTX and FVTX.

\end{enumerate}

FVTX-MuTr track mismatches can occur if MuTr tracks are projected onto a 
region where the FVTX is not active. These MuTr tracks tend to 
incorrectly associate with tracks in neighboring live regions which can 
distort the \dcar distributions. Fiducial cuts were applied to remove 
MuTr track projections onto nonactive $\phi$ regions in the FVTX and 
their edges to minimize these mismatches. Edge effects are further 
reduced by requiring the azimuthal residual between the FVTX and MuTr 
track projections is no larger than 100 mrad (corresponding to the MuTr 
azimuthal projection resolution).

Once the FVTX-MuTr association passes the matching criteria, the tracks 
are combined providing a momentum vector and projection at the vertex 
plane. The combined track has precise momentum and azimuthal direction 
mostly from the MuTr track information, and FVTX track information 
dominates the radial direction precision.

\begin{figure}[ht]
\includegraphics[width=1.0\linewidth]{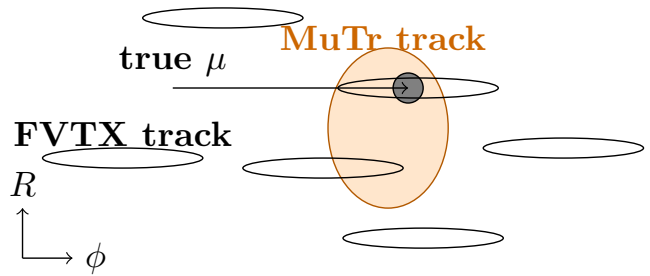}
	\caption{\label{fig:FVTX_MuTr_matching} MuTr (filled ellipse) 
and FVTX track (open ellipses) projection uncertainty areas at one of 
the FVTX planes along with the true particle (filled circle).}
\end{figure}

We determined that 65\% of MuTr tracks in Cu+Au collisions find more than one 
potential FVTX track association passing the $\chi^2$ quality and 
the matching criteria described above (Fig.~\ref{fig:FVTX_MuTr_matching}), 
whereas the probability of having more than one association is about 
15\% in \pp collisions. Because of this small fraction, the \pp analysis 
uses only the best matching. The multiple associations are pronounced in 
central Cu+Au events because of the large FVTX track density, and for 
MuTr tracks with small momentum where projection uncertainties are 
larger because of the estimated multiple scattering in the absorber. In 
\btojpsi$\rightarrow \mu \mu$ events the best $\chi^2$ association has a 
chance to happen between a MuTr track from the candidate muon and an 
FVTX track from an underlying event particle. In this situation the 
measured vertex displacement will correspond to a background particle 
and not the $B$-meson decay. In the Cu+Au analysis all FVTX-MuTr 
associations passing matching criteria (not just the best) are used in 
the \dcar distributions in order to extract the correct FVTX and MuTr 
track associations.  The contribution of mismatches to the track 
associations is then determined from event mixing, where MuTr tracks 
from one collision event are mixed with FVTX tracks from another event. 
Collisions are categorized in 10 FVTX track multiplicity and 200 
$z$-vertex classes.  The number of classes was chosen as the minimum 
number for which there was no observed change in the \dcar distributions 
of simulations embedded in real data.  MuTr tracks are mixed only with 
FVTX tracks belonging to the same collision event category. DCA 
distributions from mismatches are normalized by the relative FVTX track 
densities between same event and mixed event associations

\begin{equation}
	\label{eq:norm_mismatch}
	{\rm Norm}_{\rm mis} = \frac{\textrm{N FVTX tracks in same event}}{\textrm{N FVTX tracks in mixed events}}.
\end{equation}

\noindent The number of mixed event FVTX-MuTr associations is 
arbitrarily chosen to be five times larger than in same events in order 
to reduce background statistical fluctuations. The normalization 
\ref{eq:norm_mismatch} was tested with entire 
{\sc pythia{\footnotesize 8}}+{\sc geant{\footnotesize 4}} 
events containing prompt \jpsi and \btojpsi embedded in real data, where the 
\dcar distributions, after subtracting the normalized, event-mixed 
mismatch tracks, show excellent agreement with \dcar distributions not 
embedded in real data.

\begin{figure*}[tbh]
\includegraphics[width=0.34\linewidth]{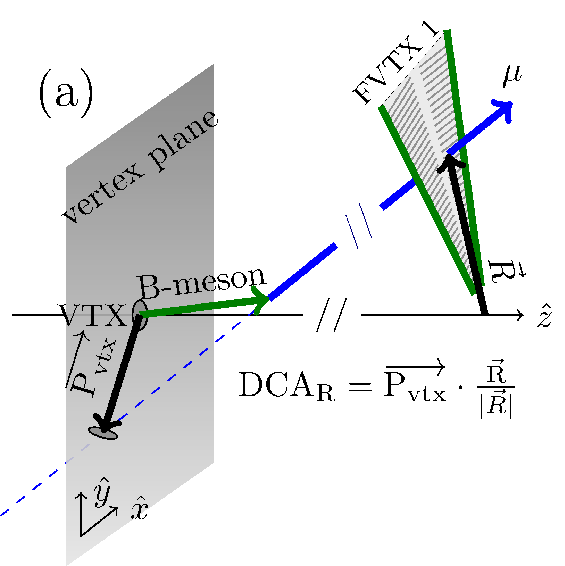}
\includegraphics[width=0.30\linewidth]{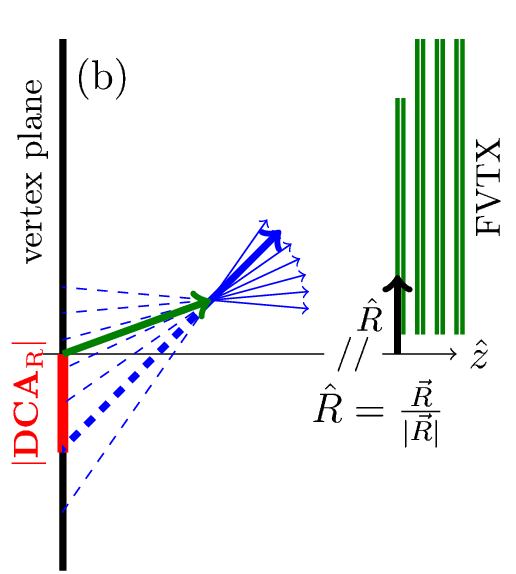}
\includegraphics[width=0.34\linewidth]{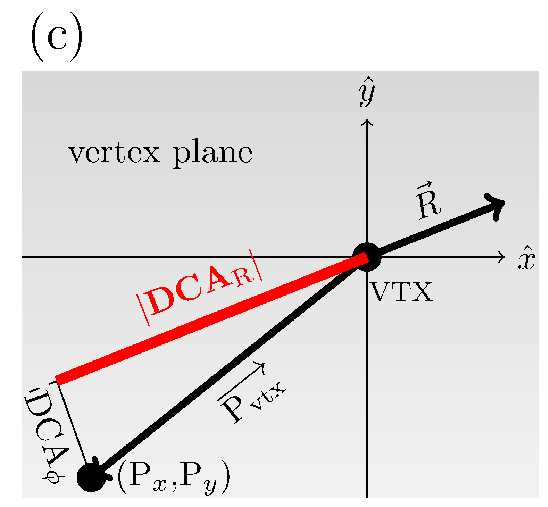}
\caption{\label{fig:DCA} Visual definition of the displaced vertex \dcar 
used in this analysis: (a) 3D, (b) $r-z$ and (c) $x-y$ views. The 
collision point VTX is at the origin in this view for simplicity.}
\end{figure*}

\subsection{Primary vertex determination.}
\label{sec:primary_vertex}

The \pp (Cu+Au) collision events are distributed in a volume, $\sigma_x 
\times \sigma_y \times \sigma_z$, of approximately $130~\mu m \times 
130~\mu m \times 40~{\rm cm}$ ($90~\mu m \times 90~\mu m \times 40~{\rm 
cm}$) centered within the PHENIX detector. Collisions within $z=\pm 10$ 
cm produce charged particle tracks in the nominal VTX and FVTX 
acceptance where the tracking can be utilized to precisely determine the 
collision point. The primary vertex location is found by the 
minimization of the squared impact parameter for the collection of 
reconstructed charged particle tracks found in the VTX and the FVTX. In 
Cu+Au, bias created by tracks with off-vertex decay points is negligible 
due to the large number of particles created in the collisions of large 
nuclei and limits on the impact parameter cluster sizes used during the 
fitting procedure. In Cu+Au, the final primary vertex position is 
determined by a three dimensional minimization on an event-by-event 
basis. The majority of the position information in the minimization 
comes from tracklets within the VTX layers due to this detector's fine 
azimuthal segmentation of 50 $\mu$m, short projection length of 
$\sim2.6$ cm, and typically large impact angles between reconstructed 
tracks. The event multiplicity is the key factor driving variance in the 
vertex resolution

\begin{equation}
\label{eq:vtx_error}
	\sigma_{\rm VTX} = \sqrt{\sigma_{\rm VTX~x}^2+\sigma_{\rm VTX~y}^2+\sigma_{\rm VTX~z}^2},
\end{equation}

\noindent where the $\sigma_{\rm VTX~x}$, $\sigma_{\rm VTX~y}$, and 
$\sigma_{\rm VTX~z}$ vertex uncertainties are obtained event-by-event 
from the vertex minimization procedure. The vertex uncertainty is 
determined from independent measurements in north, south, east and west 
parts of the VTX and FVTX detector. A check of these uncertainties is 
performed with detector simulation, using {\sc hijing}~\cite{Wang:1991hta} and 
{\sc pythia{\footnotesize 8}} generated events. The $\sigma_{\rm VTX}$ 
vertex resolution varies between 30 $\mu m$ (central collisions) and 200 
$\mu m$ (peripheral collisions) in Cu+Au collisions. Due to the low 
multiplicity in \pp collisions, the average $x$ and $y$ beam position in 
the transverse plane at the interaction point over a short data taking 
period ($\sim$90 min) are used in place of an event-by-event 
determination. Therefore, the $x$ and $y$ beam position resolutions 
correspond to the spread of the beam position (130 $\mu$m). The 
$z$-position is still determined event-by-event with typical resolution 
of 200 $\mu$m. A summary of all vertex requirements for an event to be 
selected is listed in Table \ref{tab:vertex_cuts}.

\begin{table}[ht]
	\caption{\label{tab:vertex_cuts} List of vertex position requirements for event selection.}
    \centering
    \begin{tabular}{lc}
    	requirement & cut \\\hline
        $z$ vertex position range & $|z|<$10 cm\\
        vertex resolution in \pp & $\sigma_{Z}<$500 $\mu$m \\
        vertex resolution in Cu+Au& $\sqrt{\sigma_{X}^2+\sigma_Y^2+\sigma_Z^2}<$200 $\mu$m\\\hline
    \end{tabular}
\end{table}

\subsection{Distance of closest approach measurement.}
\label{sec:DCA}

The combined FVTX+MuTr track is projected to a plane perpendicular to 
the $z$-axis and placed at the measured $z$-position of the collision 
point. The displaced vertex vector $\vec{\rm P}_{\rm vtx}$ is defined by 
the track projection $(x_{\rm track},y_{\rm track},z_{\rm vtx})$ and the 
collision point $(x_{\rm vtx},y_{\rm vtx},z_{\rm vtx})$ as illustrated 
in Fig. \ref{fig:DCA}a and \ref{fig:DCA}c.

This projection is not precise in the $\phi$ direction because of the 
coarse segmentation of the FVTX in this direction, so the radial 
projection is used in this analysis, where the FVTX has the best 
segmentation. In Fig. \ref{fig:DCA} the radial axis $\vec{R}$ is defined 
by the track projection at the first FVTX station $\left(x_{\rm 
FVTX1},y_{\rm FVTX1},z_{\rm FVTX1}\right)$ and the $z$-axis 
$\left(0,0,z_{\rm FVTX1}\right)$. The precise displacement vertex, which 
we call the radial distance of closest approach at the vertex plane 
\dcar is a projection of the vector $\vec{\rm P}_{\rm vtx}$ on the 
radial direction $\vec{R}$

\begin{equation}
\dcar = \vec{\rm P}_{\rm vtx} \cdot \frac{\vec{R}}{|\vec{R}|}.
\end{equation}

\begin{figure}[!ht]
	\centering
	\includegraphics[width=1.0\linewidth]{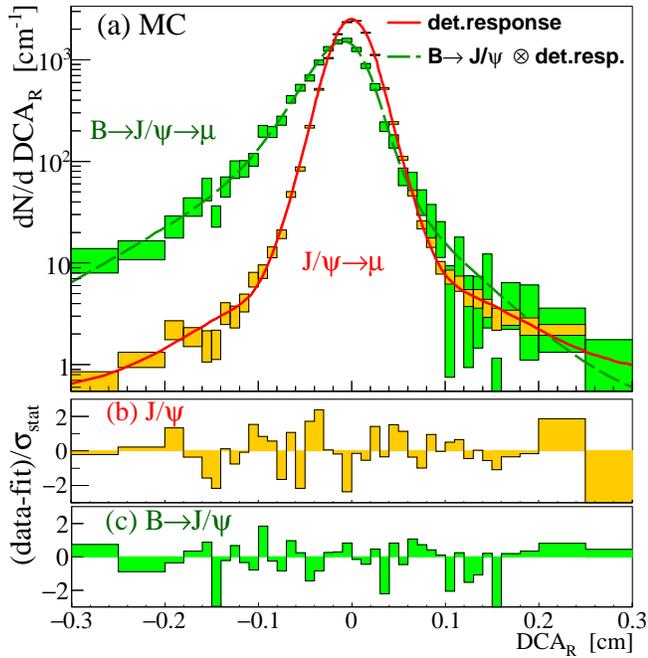}
\caption{\label{fig:mc_samples} (a) \dcar distribution of 
{\sc pythia{\footnotesize 8}}+{\sc geant{\footnotesize 4}} simulated 
prompt \jpsi and \btojpsi samples as described in 
Sec.~\ref{sec:MC_setup} (boxes) along with fitted functions 
(\ref{eq:detres}) and Eq.~(\ref{eq:decay_B}), respectively. Boxes 
represent the bin widths and statistical uncertainties in simulation. 
(b,c) Pulls between data points from simulation and fitted functions.}
\end{figure}

Figure \ref{fig:DCA}b shows a two-dimensional projection on the $r-z$ 
plane of the track. The extension of the \dcar distribution depends on 
the decay length of the parent particle and the rapidity difference 
between the parent particle and the muon. Figure \ref{fig:mc_samples} 
shows the shape of \dcar distributions of muons from simulated \btojpsi 
and prompt \jpsi decays thrown in the detector response simulation. 
Negative \dcar (where $\vec{\rm P}_{\rm vtx}$ and $\vec{R}$ have 
opposite directions) is larger for long range decays, producing an 
asymmetric \dcar distributions for muons from heavy flavor decays which 
facilitates their separation from other contributions from prompt 
particles during the fitting procedure.

\subsection{Backgrounds in the \dcar distributions.}
\label{sec:backgrounds}

\begin{figure*}[!htb]
  \includegraphics[width=0.49\linewidth]{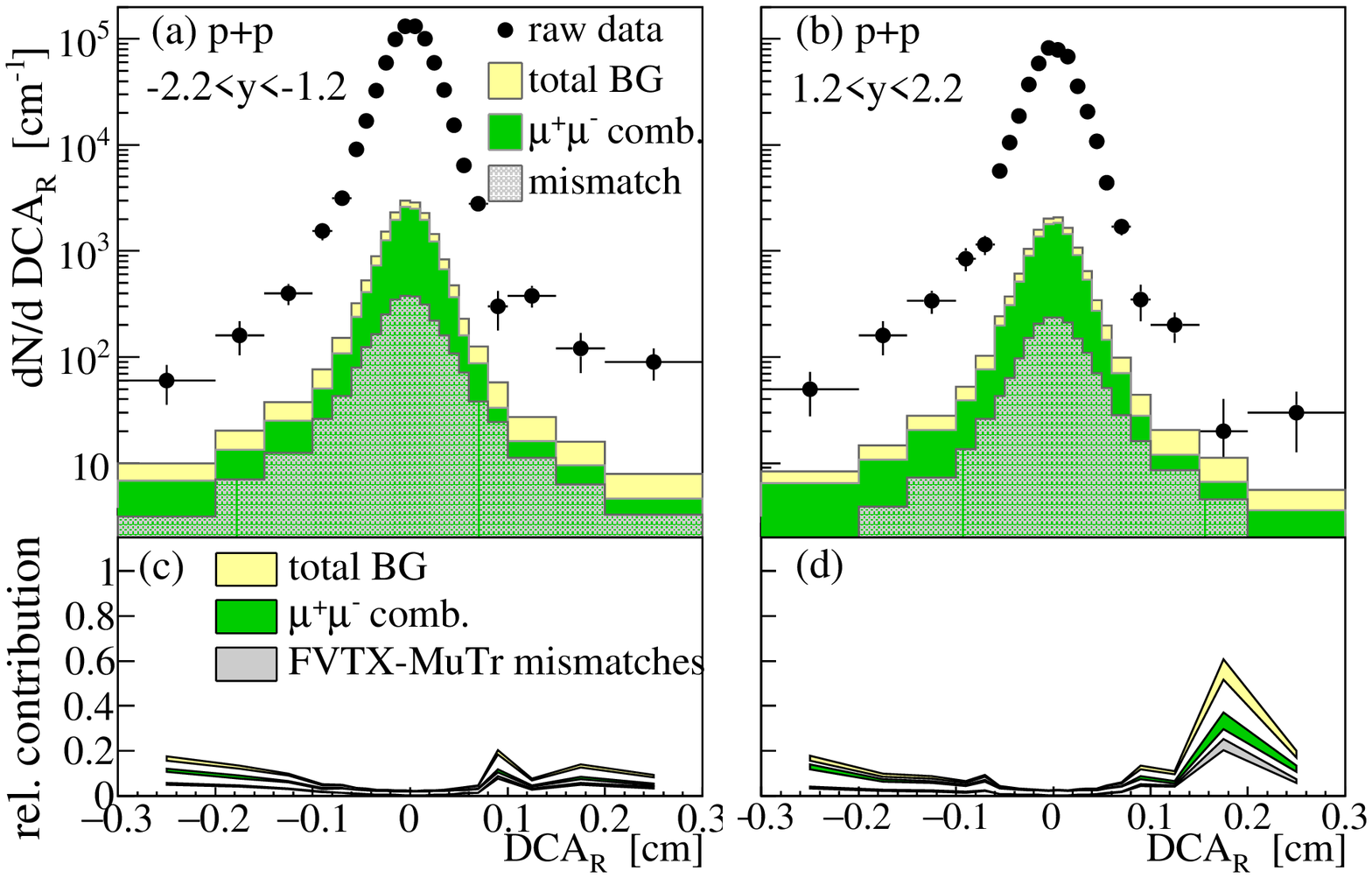}
  \includegraphics[width=0.49\linewidth]{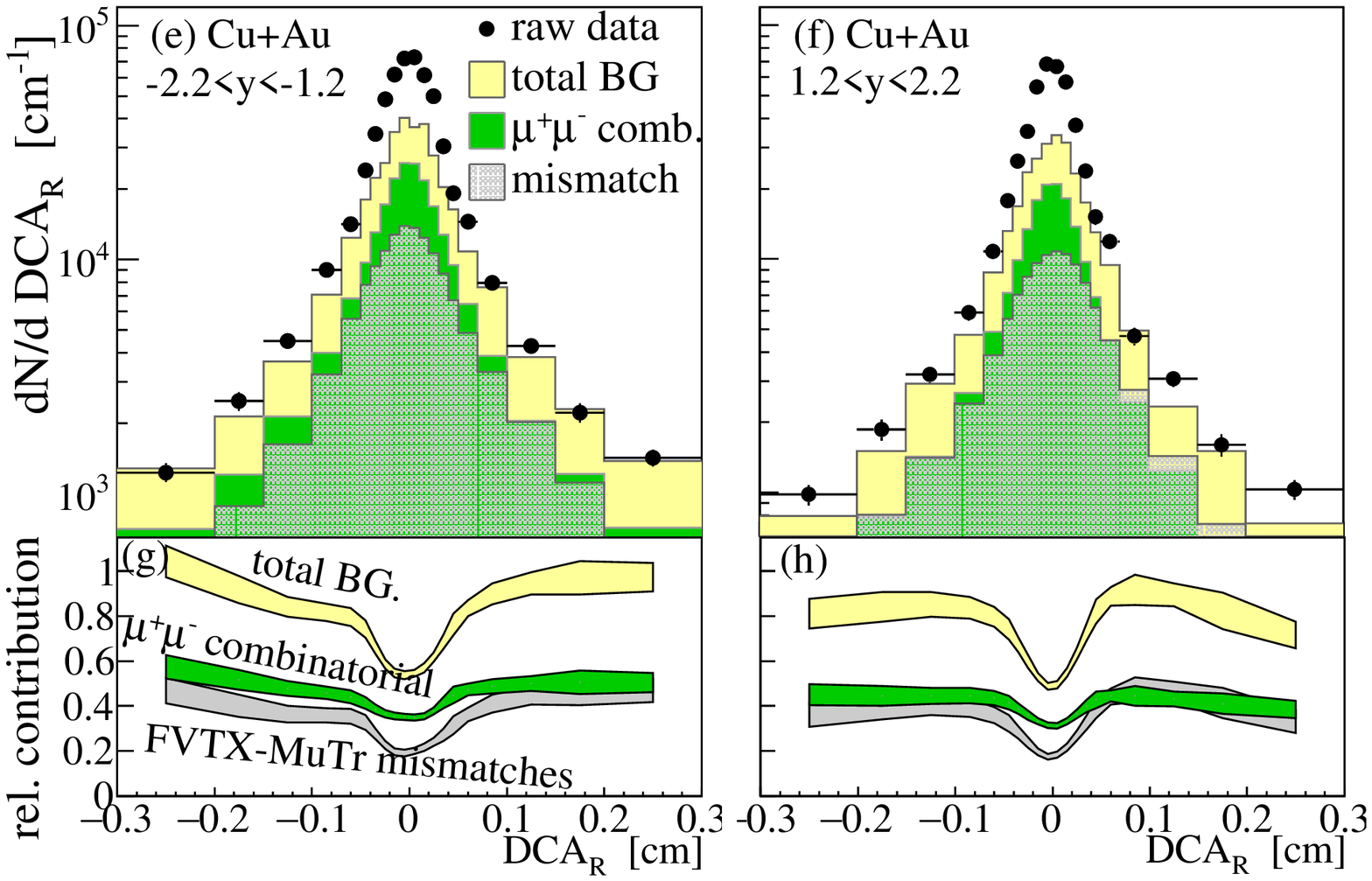}
\caption{\label{fig:dcar_backgrounds} (a,b,e,f) Uncorrelated
background contributions to the \dcar distributions from \pp (a,b) and
Cu+Au (e,f) data. (c,d,g,h) Background contribution relative to total
yield. Bands in the relative contributions correspond to statistical
uncertainties.}
\end{figure*}

The \dcar distributions comprise MuTr tracks, from dimuons in 
the \jpsi mass region, which are associated with one or more FVTX 
tracks. Each FVTX+MuTr association counts in the \dcar distributions. 
There are two significant sources of uncorrelated background in the 
\dcar distributions: 1) combinatorial background described in Section 
\ref{sec:jpsi_selection} in the \jpsi mass region, corresponding to 
dimuons which are not from \jpsi decays; and 2) FVTX-MuTr mismatches 
described in Section \ref{sec:FVTX_MuTr_association} corresponding to 
MuTr tracks from \jpsi decays but associated to the wrong FVTX track. 
Figure \ref{fig:dcar_backgrounds} shows the contribution of these 
backgrounds, extracted using the techniques described above, in the raw 
\dcar distribution obtained in both arms. The combinatorial background 
distribution is obtained from mixed event dimuons, and the mismatch 
contribution is determined from event mixed MuTr-FVTX associations. 
Statistical fluctuations are reduced by obtaining five times as many 
mixed event pairs and misassociations than in same event backgrounds. 
The normalization of these two distributions was explained in Sections 
\ref{sec:jpsi_selection} and \ref{sec:FVTX_MuTr_association}. According 
to the distributions shown in Fig.~\ref{fig:dcar_backgrounds}, most 
background contributions come from prompt particles, but the relative 
background contribution changes at large $|\dcar|$ where fake or bad 
quality tracks and muons from light hadron decays are more significant.

The \dcar line shape of correlated background contributions from \cc and 
\bb is obtained from simulation and discussed in Sec. 
\ref{sec:decays_mc}.

\section{Simulation.}
\label{sec:MC_setup}

In this Section we describe how the \dcar line shape of each dimuon 
source in the \jpsi mass region is obtained from simulation.

\subsection{Detector response of prompt particles.}
\label{sec:det_response}

The detector response to prompt decays, such as muons from prompt \jpsi, 
is estimated using prompt hadrons (pions, kaons and protons) generated 
by the {\sc pythia{\footnotesize 8}} event generator in MB mode and a 
{\sc geant{\footnotesize 4}}-based detector simulation package 
\cite{Agostinelli2003250}. Several dead channel configurations are used 
to account for run-by-run detector acceptance fluctuations. This 
procedure was found to be crucial to simulate remaining edge effects 
(Sec.~\ref{sec:FVTX_MuTr_association}) and degradation of FVTX-VTX track 
quality which can potentially produce long tails in the measured \dcar 
distributions. The simulated {\sc geant{\footnotesize 4}} signals from 
all generated hadrons in a single {\sc pythia{\footnotesize 8}} event 
are embedded in real data to account for occupancy and accidental 
hit-track-association effects in the \dcar measurement. All particles 
generated by the detector simulation are shifted by 
$\func{\delta x, \delta y, \delta z}$, where each of the coordinates is 
a common Gaussian random number centered at the vertex position measured 
in the real data event and with a width determined from the vertex 
uncertainty of the real event the simulated particles are embedded in. 
Event reconstruction, including the simulated hits and the entire real 
data signal in the event, is then performed in the same manner as with 
real data.

\begin{figure*}[!ht]
        \includegraphics[width=0.48\linewidth]{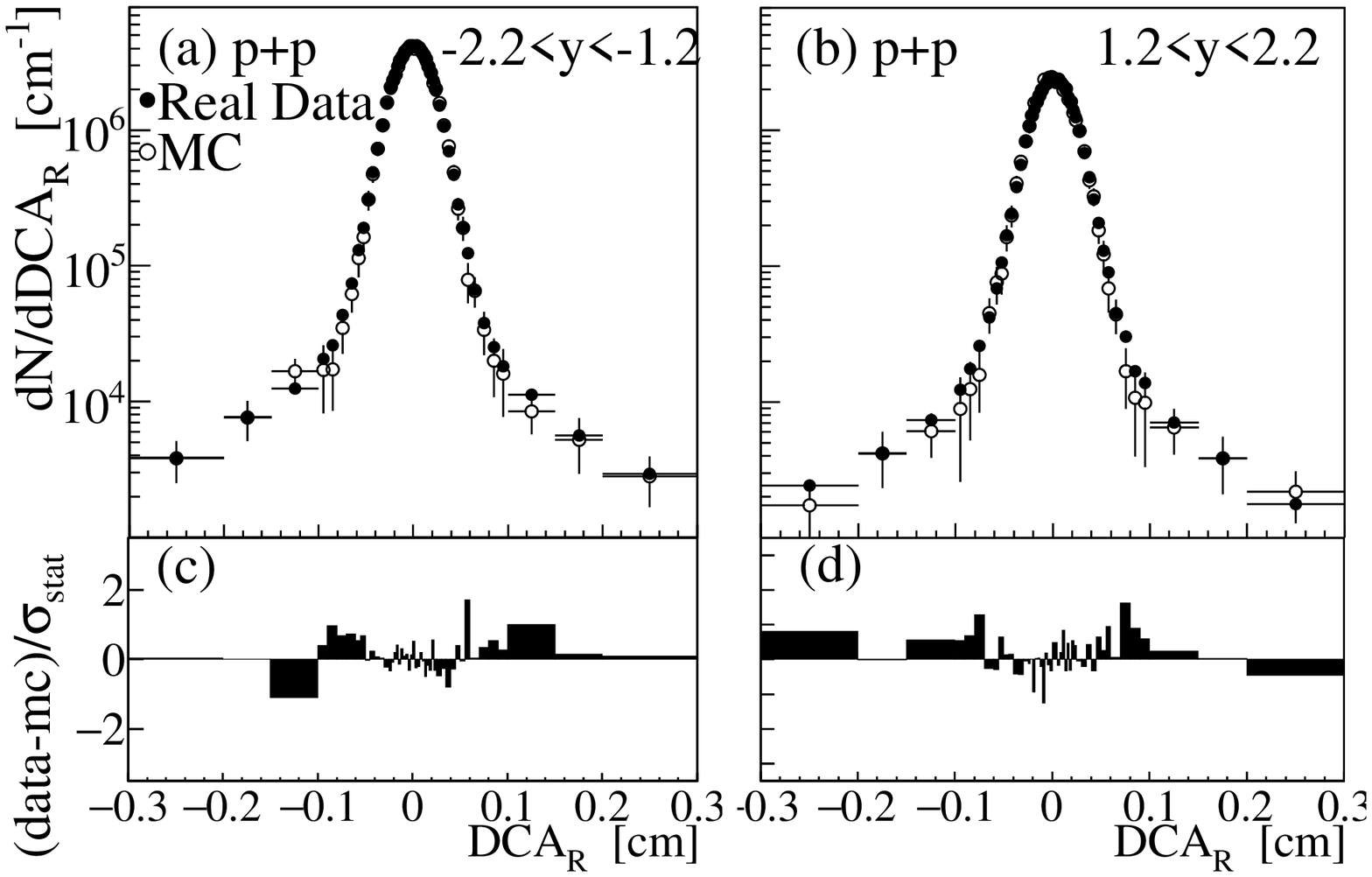}
        \includegraphics[width=0.48\linewidth]{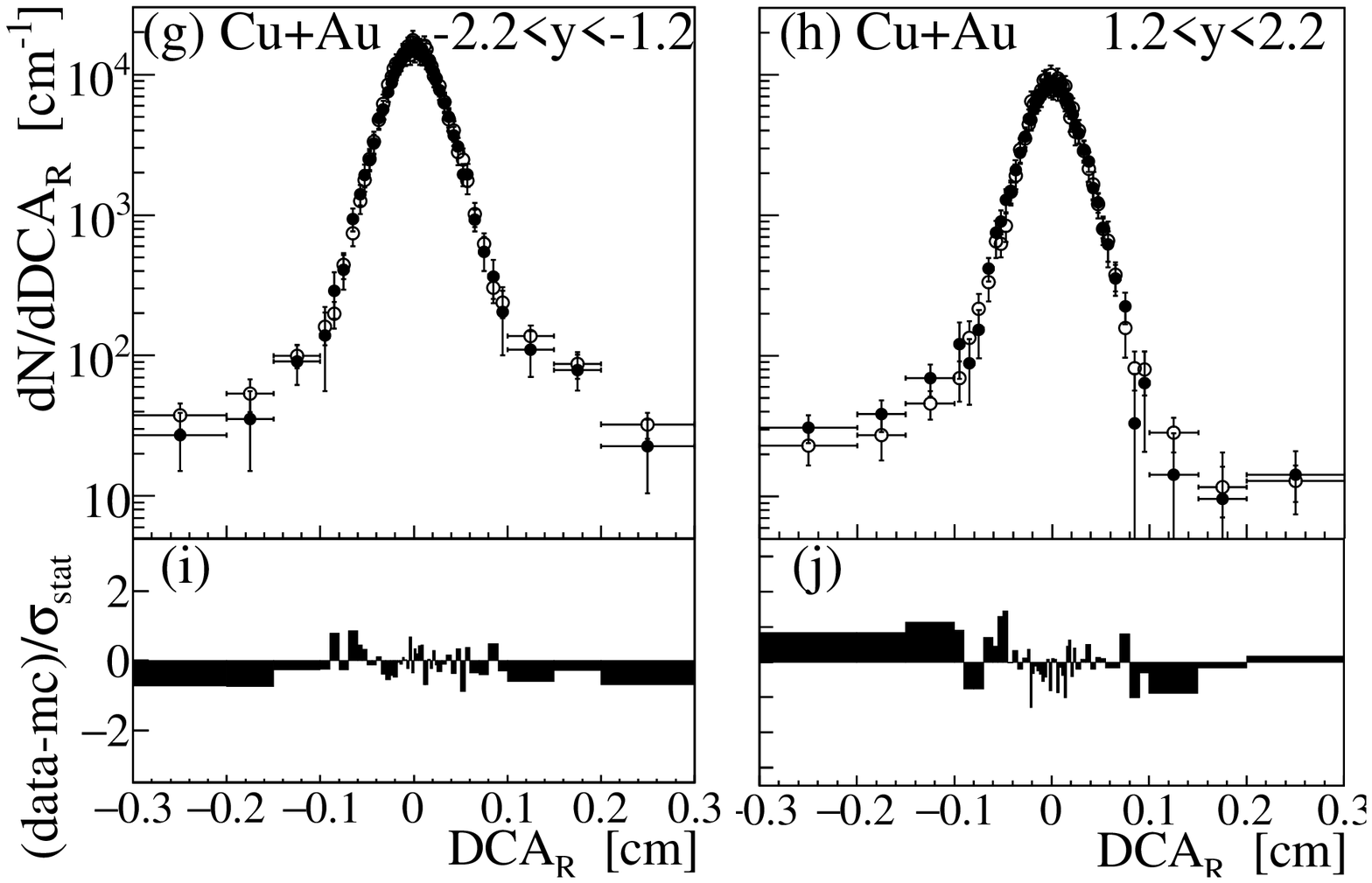}
        \includegraphics[width=0.508\linewidth]{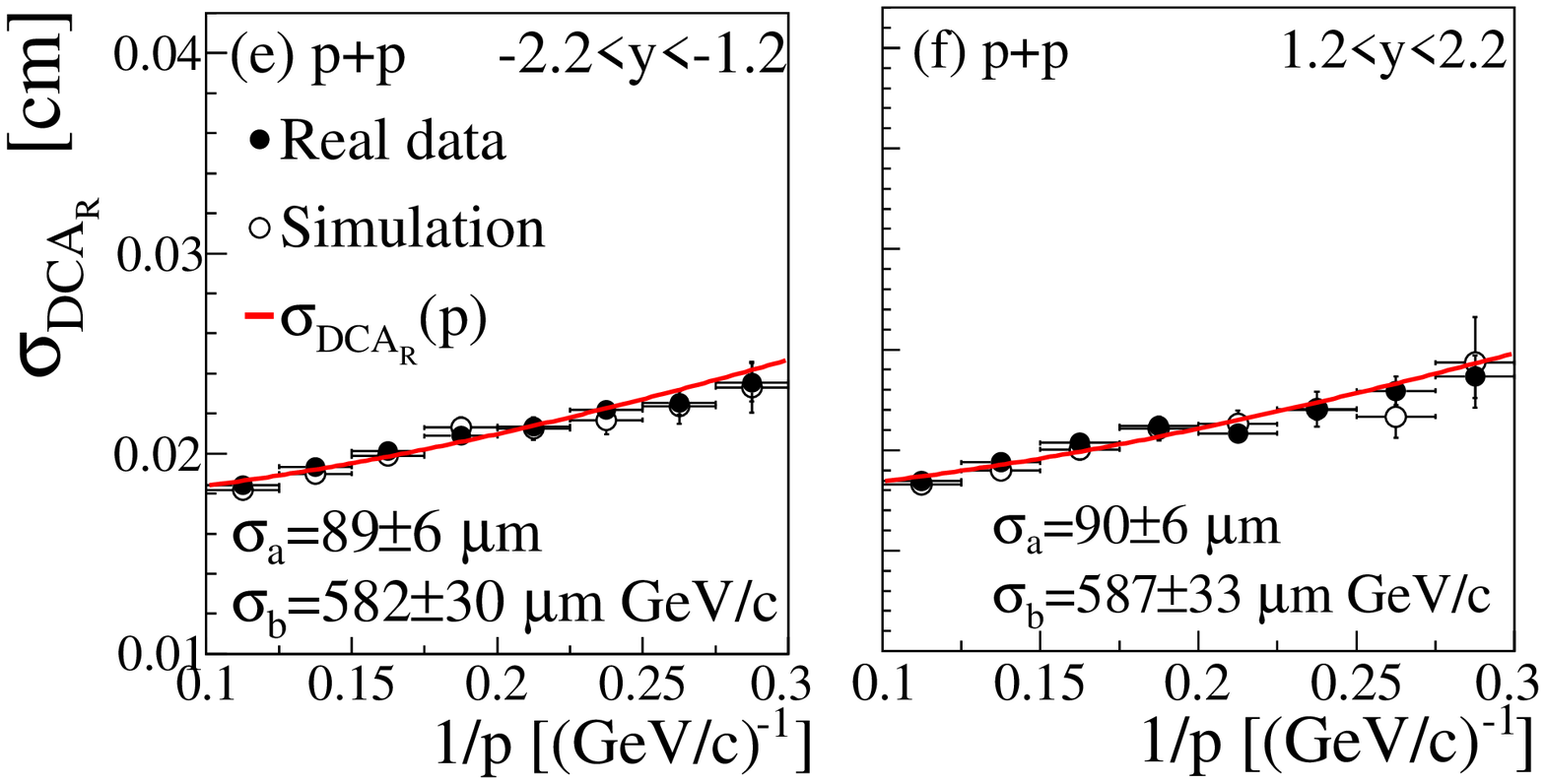}
        \includegraphics[width=0.48\linewidth]{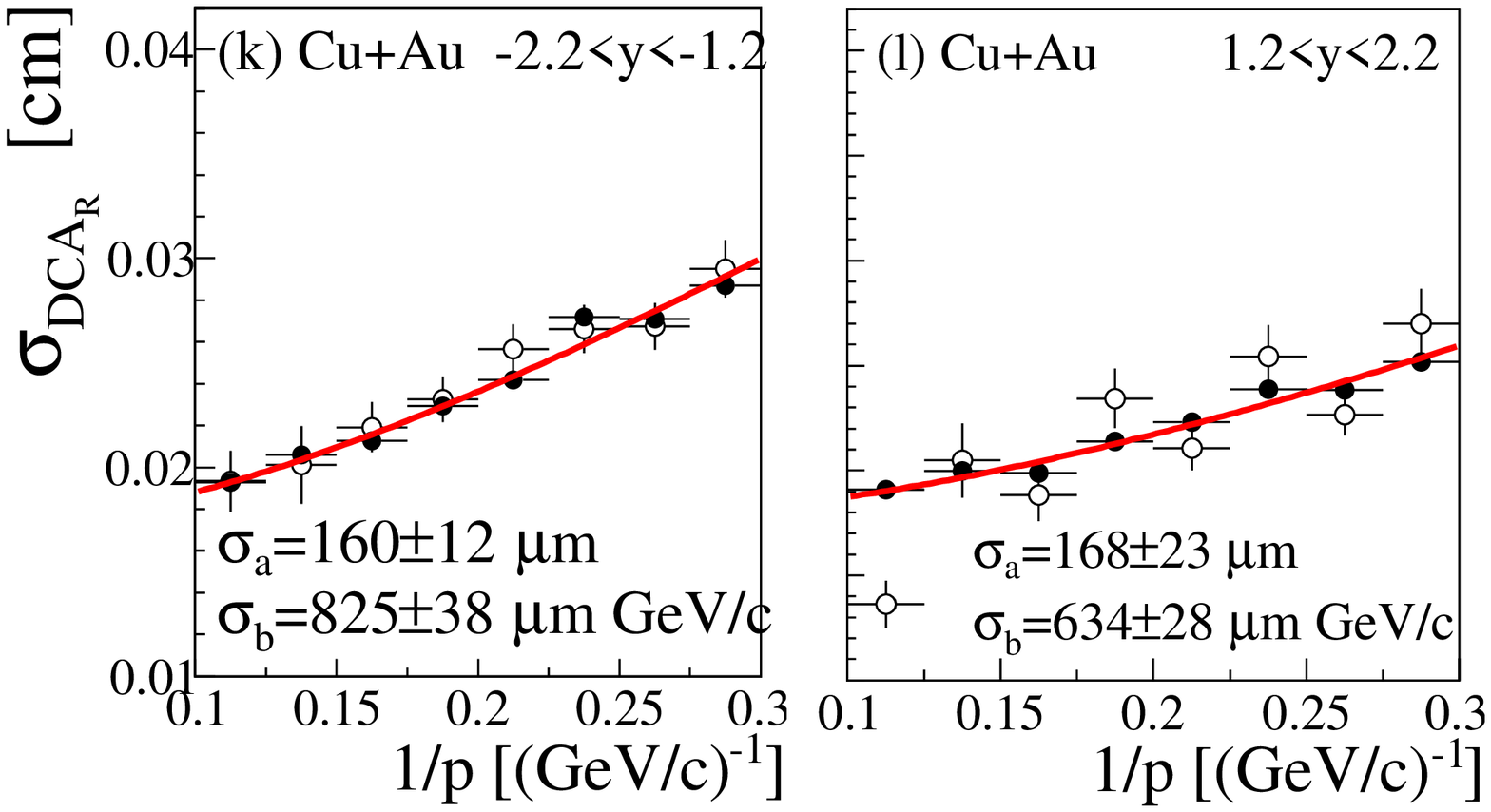}
\caption{\label{fig:gap2} \dcar distribution of light hadrons in (a,b) 
\pp and (g,h) Cu+Au real data after (closed circles) FVTX-MuTr mismatch 
subtraction.  (open circles) momentum weighted simulation. (c,d,i,j) 
Pulls between real data and simulation. (e,f,k,l) Momentum dependence of 
\dcar$\in$[-0.04,0.04] cm width in real and simulated hadron data along 
with fitted function (\ref{eq:dcar_resolution}).} 
\end{figure*}

Additional smearing and offsets are needed in the simulated \dcar in 
order to account for irreducible detector misalignments, additional 
smearing of the primary vertex resolution that is not accounted for in 
the simulations, and any missing materials in the simulation that could 
be a multiple scattering source between the interaction region and the 
last FVTX plane. The measured \dcar resolution depends on 1) the vertex 
uncertainty $\sigma_{\rm VTX}$ defined in Eq.~(\ref{eq:vtx_error}); 2) a 
momentum independent parameter $\sigma_a$ which represents the sum of 
the detector resolution contribution to the \dcar resolution, additional 
vertex uncertainty which might not be captured by $\sigma_{\rm VTX}$, 
residual detector misalignments and possible beam line tilt variations 
during the run; and 3) a multiple scattering term $\sigma_b$ in the VTX 
and FVTX material which introduces the momentum dependence in the 
resolution:

\begin{equation}
\label{eq:dcar_resolution}
	\sigma_{\rm DCA_{R}}\left(p,\sigma_{\rm vtx}\right) = \sqrt{\sigma_{\rm vtx}^2 + \sigma_a^2 + \left(\sigma_b/p\right)^2}.
\end{equation}

Particles which stop at the third absorber plane in the MuID comprise
mainly prompt light hadrons and $\sim$7\% contribution from 
hadron$\rightarrow\mu$ decays according to 
{\sc pythia{\footnotesize 8}}+{\sc geant{\footnotesize 4}} 
simulation.  This sample provides a relatively clean selection of prompt 
particles which can be used to compare \dcar spectra between simulation 
and real data. The \dcar distributions from the measured prompt hadrons 
are compared to simulated hadrons, and small smearing corrections are 
added to the simulated spectra until they match the real data spectra.
 
For real data and simulations, the $\sigma_{\rm VTX}$ value is set 
according to the event-by-event vertex resolution that is provided by 
the vertex finding code. The fit parameters $\sigma_a$ and $\sigma_b$ 
are extracted from the real data distributions, and the simulated data 
has additional smearing added until the distributions produce the same 
$\sigma_a$ and $\sigma_b$ fit results. An additional smearing of 70(110) 
$\mu$m was found to be necessary in the \pp(Cu+Au) setup. After tuning 
the simulation, the \dcar distribution of simulated hadrons stopping in 
the MuID is weighted according to the momentum distribution of real data 
stopped hadrons. 

Figure~\ref{fig:gap2} shows the \dcar distribution of these prompt 
hadrons in real data compared to {\sc pythia{\footnotesize 8}} generated 
hadron events which have had tuned smearing parameters added. Figure 
\ref{fig:gap2} also shows the final fit parameters $\sigma_a$ and 
$\sigma_b$ for each of the north (positive rapidity) and south (negative 
rapidity) arms. Although the material present in the north and south 
arms is nominally identical, the momentum-dependent resolution parameter 
$\sigma_b$ is found to be different for the two arms in the Cu+Au data 
set. The reason for this is that the vertex distribution for 
BBC-triggered events in Cu+Au events was found to be highly skewed 
toward the north arm, given the collision species asymmetry. Because of 
this, FVTX tracks had on average a shorter projection to the vertex for 
tracks in the north arm than in the south arm, resulting in a smaller 
contribution to the \dcar resolution from multiple scattering in the 
north than in the south. Long tails in the \dcar distribution contain 
part of the light hadron decay contribution but are mostly dominated by 
accidental hit-track associations which are reasonably well reproduced 
by the embedded simulation in the \dcar range used in this analysis.

\subsection{Prompt \jpsi and heavy flavor decay simulations.}
\label{sec:decays_mc}

Prompt \jpsi and \btojpsi decays are generated using the 
{\sc pythia{\footnotesize 8}} event generator; open heavy flavor from 
\cc and \bb processes are produced with leading order processes (gluon 
fusion) and the CT10 parton density distribution functions 
\cite{Lai:2010vv}. All particles generated in the \btojpsi or prompt 
\jpsi event are used as input to the {\sc geant{\footnotesize 4}} 
detector simulation and are embedded in real data. Generated events have 
the \dcar values smeared and offset according to the parameters obtained 
for light hadrons as described in Section \ref{sec:det_response}. 
Simulated events are also weighted according to the momentum 
distribution of background-subtracted (combinatorial dimuon and 
FVTX-MuTr mismatch) muons from real data dimuons. Figure 
\ref{fig:mc_samples} shows the \dcar distributions obtained from these 
prompt \jpsi and \btojpsi simulations. Several hypotheses of \btojpsi 
nuclear modifications are considered in the momentum weighting for 
systematic uncertainty evaluations and described in Section 
\ref{sec:weighting}. Point-to-point statistical fluctuations in the 
generated \dcar distributions are minimized by using fitted analytical 
functions for the final \dcar fits. The \dcar distributions obtained 
from prompt \jpsi simulation are well described by a three-Gaussian 
($G_1$, $G_2$ and $G_3$) function with a detector offset and resolutions 
$\sigma_1$, $\sigma_2$ and $\sigma_3$, and regulated by $f_1+f_2+f_3=1$:

\begin{eqnarray}
  \label{eq:detres}
  \frac{{\rm det.resp}\func{\dcar}}{N} &=& f_1 G_1(\dcar; {\rm offset}, \sigma_1)\\\nonumber
    &+& f_2 G_2\func{\dcar; (dca0_2-{\rm offset}), \sigma_2} \\\nonumber
    &+& f_3 G_3\func{\dcar; (dca0_3-{\rm offset}), \sigma_3}.
\end{eqnarray}

\noindent The following function defines the prompt \jpsi \dcar line 
shape of prompt \jpsi decays

\begin{equation}
	{\rm decay}_{J/\psi}\func{\dcar} = \delta(0) \otimes {\rm det.resp}\func{\dcar}.
\end{equation}

The true \dcar distribution for heavy flavor decays is described by a 
set of three exponential functions:

\begin{widetext}
\begin{equation}
{\rm decay}^{\rm true}(\textrm{DCA}_{\rm R}) = \left\{\begin{array}{rl} 
{\rm fd}_{\rm 1} e^{\frac{-\textrm{DCA}_{\rm R}}{\lambda_{\rm l1}}} 
+ {\rm fd}_{\rm 2} e^{\frac{-\textrm{DCA}_{\rm R}}{\lambda_{\rm l2}}} 
& \textrm{DCA}_{\rm R}<0 \\
(1-{\rm fd}_{\rm 1}-{\rm fd}_{\rm 2}) e^{\frac{-\textrm{DCA}_{\rm R}}{\lambda_{\rm r}}} 
& \textrm{DCA}_{\rm R}>0
	\end{array}\right.
\label{eq:decay_B_true}
\end{equation}
\end{widetext}

\noindent which must be convoluted with the detector response function 
(\ref{eq:detres}) extracted from the prompt \jpsi fit in order to obtain 
the measured \dcar distribution:

\begin{equation}
{\rm decay}(\textrm{DCA}_{\rm R}) 
= {\rm decay}^{\rm true}(\textrm{DCA}_{\rm R}) \otimes {\rm det. resp}\func{\textrm{DCA}_{\rm R}}.
	\label{eq:decay_B}
\end{equation}

\begin{figure}[!hbt]
	\includegraphics[width=1.0\linewidth]{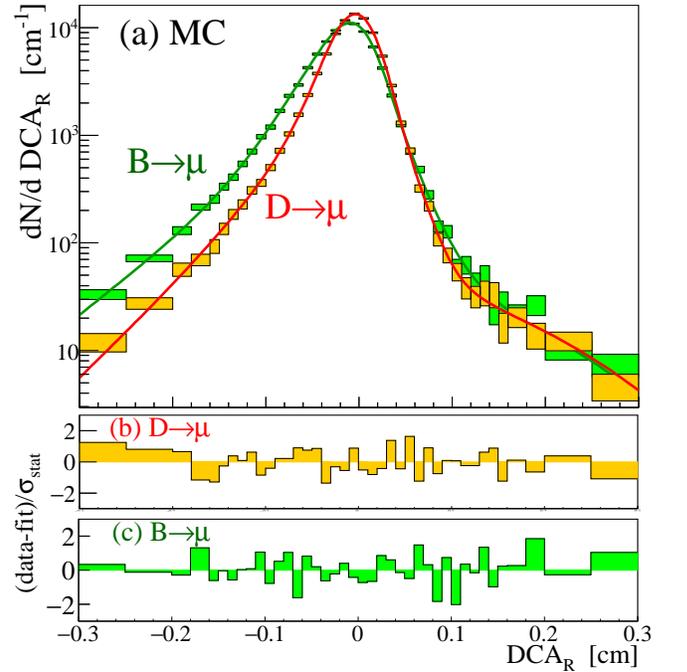}
\caption{\label{fig:corrbg_dcar} (a) Simulated correlated background 
components, in the dimuon mass $\in[2.9,3.5]$ \gevcsq, after momentum 
weighting and vertex uncertainty along with fitted function 
(\ref{eq:decay_B}). Boxes represent the bin widths and statistical 
uncertainties in simulation. (b,c) Pulls between simulated data points 
and fitted functions.}
\end{figure}

\noindent The dashed line in Fig. \ref{fig:mc_samples} shows function 
(\ref{eq:decay_B}) fitted to the simulated \btojpsi decays. Function 
(\ref{eq:decay_B}) is also used to fit the simulated correlated 
background components \cc and \bb as seen in Fig. \ref{fig:corrbg_dcar}. 
The heavy flavor decay functions used in the final fit are:

\begin{eqnarray}
	{\rm decay}_D\func{\dcar} &=& {\rm decay}\func{\dcar; 
\overrightarrow{\rm par}_{D\rightarrow\mu}}\\\nonumber
    {\rm decay}_B\func{\dcar} &=& {\rm decay}\func{\dcar; 
\overrightarrow{\rm par}_{B\rightarrow\mu}}\\\nonumber
    {\rm decay}_{B{\rightarrow}J\psi}\func{\dcar} 
&=& {\rm decay}\func{\dcar; \overrightarrow{\rm par}_{B{\rightarrow}J/\psi\rightarrow\mu}}\nonumber
\end{eqnarray}

\noindent where $\overrightarrow{\rm par}_{D\rightarrow\mu}$, 
$\overrightarrow{\rm par}_{B\rightarrow\mu}$ and 
$\overrightarrow{\rm par}_{B{\rightarrow}J/\psi\rightarrow\mu}$ are the 
parameters in Eq.~(\ref{eq:decay_B_true}) fitted to the corresponding 
heavy flavor decay simulations.

\section{Fitting Procedure.}
\label{sec:fitting_procedure}

The final fitting function used for the real data muon \dcar 
distribution is

\begin{eqnarray}
\label{eq:final_fit}
	f_{\mu}\func{\dcar} 
&=& {\rm BG}_{\rm uncor}\func{\dcar} + 
{\rm BG}_{\rm corr}\func{\dcar} \\\nonumber
&+& N_{J/\psi}\left[\func{1-F_{B{\rightarrow}/J\psi}^{\rm meas}}
{\rm decay}_{J/\psi}\func{\dcar}\right. \\\nonumber
&+& \left.{\rm F}_{B{\rightarrow}J\psi}^{\rm meas}\cdot{\rm decay}_{B{\rightarrow}J\psi}\func{\dcar}\right] 
\end{eqnarray}

\noindent where BG$_{\rm uncorr}$ includes the normalized dimuon 
combinatorial (Sec.~\ref{sec:jpsi_selection}) and FVTX-MuTr mismatch 
(Sec.~\ref{sec:FVTX_MuTr_association}) backgrounds. The correlated 
background BG$_{\rm corr}$ comprises \cc and \bb contributions:

\begin{eqnarray*}
	\label{eq:bg_corr}
{\rm BG}_{\rm corr}\func{\dcar} &=& N_{\rm corr} 
f_{\rm cont}[(1-{\rm f}_{b\bar{b}}){\rm decay}_D\func{\dcar} \\\nonumber
    &+& {\rm f}_{b\bar{b}}{\rm decay}_B\func{\dcar}]\nonumber.
\end{eqnarray*}

\noindent where $N_{\rm corr}$ is the number of muons after subtracting 
BG$_{\rm uncorr}$. The continuum correlated background $f_{\rm cont}$ is 
defined in Table \ref{tab:jpsi_counts}. The muon count from 
inclusive \jpsi decays is $N_{\jpsi} = (1-f_{\rm cont})N_{\rm corr}$. 
The fraction of \bb contribution in the correlated background 
f$_{b\bar{b}}$ was determined from extrapolations of previous \cc and 
\bb cross section measurements \cite{Adare:2006hc,Adare:2009ic} 
indicating a fraction f$_{b\bar{b}}$=0.32 $\pm$ 0.21 in \pp collisions 
at $\sqrt{s}=$ 200 GeV. The \bb contribution in the correlated 
background is completely undetermined in Cu+Au collisions. So we set 
f$_{b\bar{b}}$=0.5 in the standard fit and vary it in the range 
f$_{b\bar{b}}\in$[0,1] for the systematic uncertainty evaluation.

\begin{figure*}[!htb]
\begin{minipage}{0.99\linewidth}
\includegraphics[width=0.99\linewidth]{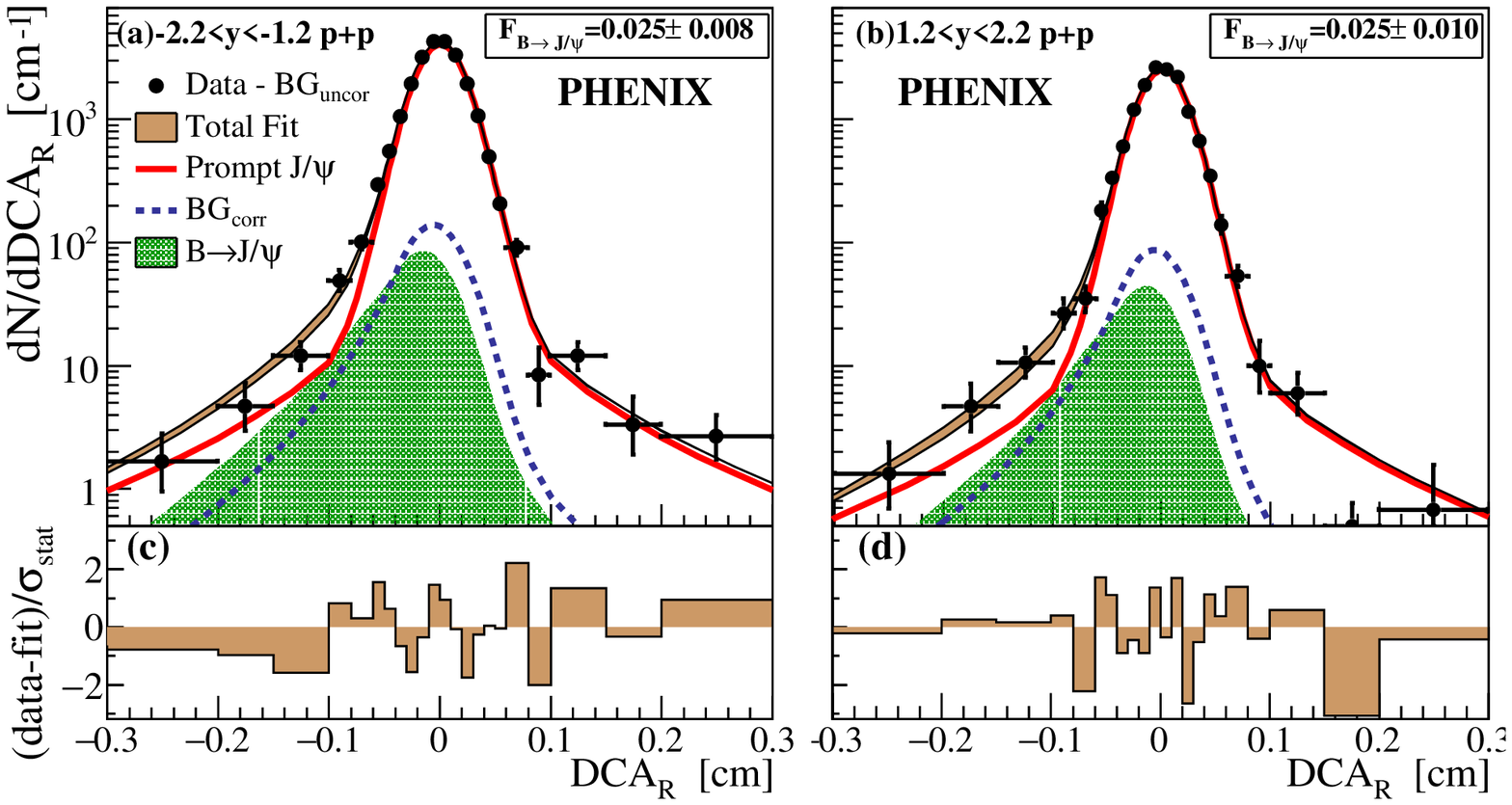}
\caption{\label{fig:finalfits_pp} (a,b) Function (\ref{eq:final_fit}) 
fitted to the \pp data in different rapidity ranges. The combinatorial 
and FVTX-MuTr mismatch backgrounds are subtracted from both data and the 
fitting function for clarity. The band around the total fit curve 
corresponds to the propagated fitting uncertainty. The resulting \bfrac 
is corrected by the relative acceptance and efficiency, and the 
evaluated uncertainty is only from the fit. (c,d) Pulls between data 
points and fitted functions.}
\end{minipage}
\begin{minipage}{0.99\linewidth}
\includegraphics[width=0.99\linewidth]{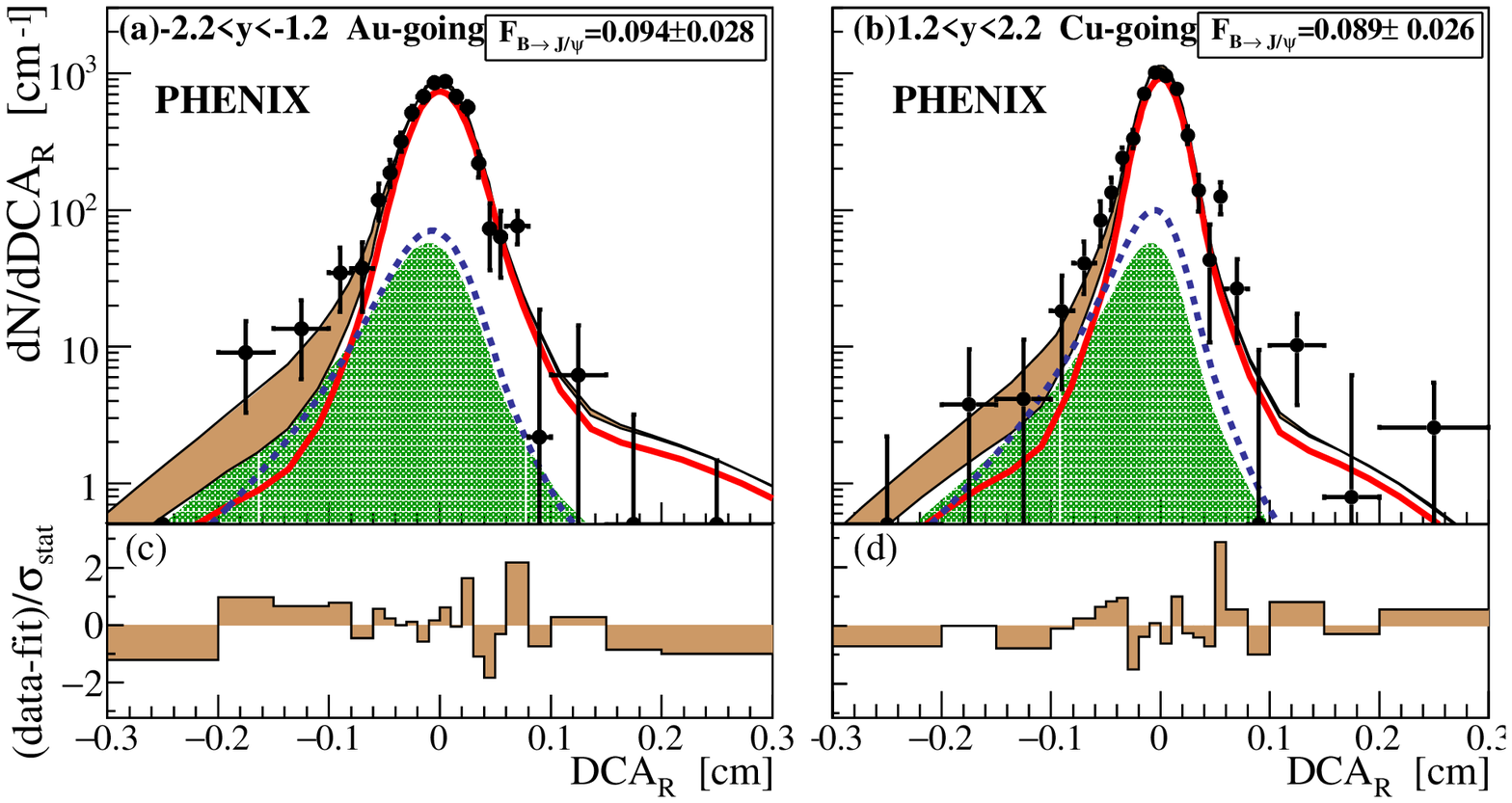}
\caption{\label{fig:finalfits_cuau}(a,b) Function (\ref{eq:final_fit}) 
fitted to the Cu+Au data in different rapidity ranges. The combinatorial 
and FVTX-MuTr mismatch backgrounds are subtracted from both data and the 
fitting function for clarity. The band around the total fit curve 
corresponds to the propagated fitting uncertainty. The resulting \bfrac 
is corrected by the relative acceptance and efficiency and the evaluated 
uncertainty is only from the fit. (c,d) Pulls between data points and 
fitted functions.}
\end{minipage}
\end{figure*}

The measured fraction of $B$-meson decays 
F$_{B{\rightarrow}J\psi}^{\rm meas}$ is the only free parameter when 
fitting function (\ref{eq:final_fit}) to the real data muon \dcar 
distribution. The true \bfrac is obtained by correcting the measured 
\bfrac with a factor given by the relative detector acceptance and 
efficiency of \jpsi from $B$-meson decays with respect to prompt \jpsi 
decays

\begin{equation}
\label{eq:acceptance_corrected_bfrac}
\frac{1}{F_{B{\rightarrow}J\psi}} = 1 + \left(\frac{1}{F_{B{\rightarrow}J\psi}^{\rm meas}}-1\right)\frac{\varepsilon_B}{\varepsilon_{J/\psi}},
\end{equation}

\noindent where $\varepsilon_B/\varepsilon_{J/\psi}\in$[0.96, 0.98] 
depending on the data set (\pp or Cu+Au) and muon arm.

The fit is performed with an unbinned extended log-likelihood method, 
where

\begin{equation}
	\label{eq:likelihood_fit}
    -\ln \mathcal{L} = \sum_{i=1}^{i=N_{\mu}} -\ln f_{\mu}\func{\dcar;\bfrac} - N_{\mu}
\end{equation}

\noindent is minimized. $N_{\mu}$ is the number of muons in the \dcar 
distribution, including the backgrounds. Figures \ref{fig:finalfits_pp} 
and \ref{fig:finalfits_cuau} show the fitted function 
(\ref{eq:final_fit}) to the \dcar distributions in \pp and Cu+Au 
collisions, respectively. The backgrounds due to combinatorial dimuons 
and FVTX-MuTr mismatches are subtracted for clarity in the figures. The 
bars in the figure show the total uncertainties of each data point.


\section{Systematic uncertainties.}
\label{sec:sys_errors}

The systematic uncertainties are determined by fitting 
$f_{\mu}\func{\dcar}$ in function (\ref{eq:final_fit}) 
several times, using random variations of the fitting parameters for 
each fit. This section lists all considered systematic uncertainty 
sources and how they affect the final results.

\subsection{List of systematic uncertainty sources.}

\subsubsection{Testing of the fitting procedure.}
\label{sec:fitting_procedure_sys}

The validity of the fitting procedure was tested by randomly generating 
\dcar distributions for backgrounds, \btojpsi, and prompt \jpsi, with 
\bfrac tested in the range [0, 0.4]. The total number of entries in the 
summed \dcar distributions were generated to match the real data 
distributions. The fit results over thousands of randomly generated 
\dcar distributions in each F$_{B{\rightarrow}J\psi}^{\rm generated}$ return 
F$_{B{\rightarrow}J\psi}^{\rm measured}$ with average bias 
$\left|{\rm F}_{B{\rightarrow}J\psi}^{\rm generated}-{\rm F}_{B{\rightarrow}J\psi}^{\rm 
measured} \right|<0.005$ in Cu+Au data and negligible in \pp data.

\subsubsection{Weighting of simulated samples.}
\label{sec:weighting}

\begin{figure}[!ht]
	\includegraphics[width=0.98\linewidth]{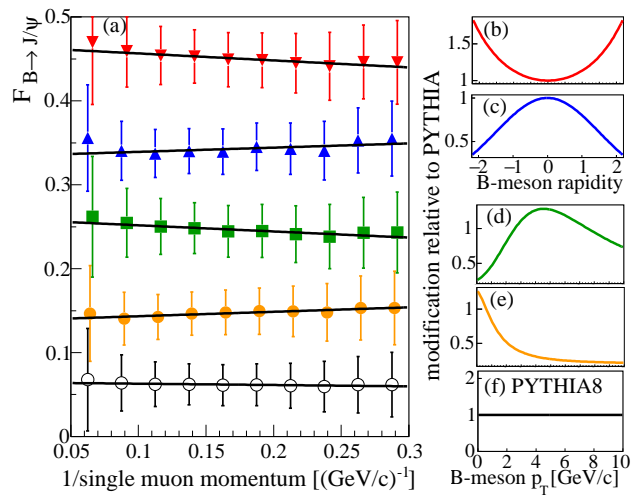}
\caption{\label{fig:bfrac_raa_hypothesis} (a) Momentum distribution of 
\bfrac from {\sc pythia{\footnotesize 8}} (open circles) and a set of 
\pt and rapidity $B$-meson yield modifications relative to prompt 
\jpsi (closed symbols) defined in panels (b,c,d,e,f). Scales on 
\bfrac are arbitrary.}
\end{figure}

All detector reconstructed simulated samples were weighted according to 
the single muon momentum distributions of real data muons from \jpsi 
decays used in the \dcar distributions. This weighting already accounts 
for the realistic momentum distribution of measured dimuons in the \jpsi 
mass region. Figure \ref{fig:bfrac_raa_hypothesis} shows how the 
momentum distribution of muons from \btojpsi decays are distributed 
relative to muons from prompt \jpsi decays in several conservative 
$B$-meson $p_T$ and rapidity yield modification hypotheses. The inverse 
momentum dependence obtained from the several hypotheses can be 
described by polynomials of degree one $f_{B{\rightarrow}J\psi}(p)$. The 
maximum deviation from the assumption used in the standard result is 
$\left|f_{B{\rightarrow}J\psi}(p)-\bfrac\right|<0.005$. This value is 
used in the systematic uncertainty determination.

\subsubsection{\dcar resolutions and offsets in simulation.}
\label{sec:det_response_sys}

A relative variation of 15\% in the detector offset, 5\% for $\sigma_a$ 
and 7\% for $\sigma_b$ are found when different fitting ranges are used 
to determine the offset and the momentum dependence of the \dcar 
detector resolution shown in Figs.~\ref{fig:gap2}c,d,g,h.  
Varying these parameters within their uncertainties produces fluctuations 
on the \bfrac result of up to 0.015.

\subsubsection{\dcar tail contribution to the detector response function.}
\label{sec:f_core_sys}

The detector response function (\ref{eq:detres}) is fitted to simulated 
and real data hadrons shown in Fig.~\ref{fig:gap2}. The third Gaussian, 
which accounts for the long range tails in the \dcar distributions, has 
a contribution which is 20\% different between real and simulated data. 
This difference can be caused by 1) modified light hadron decay in real 
data, 2) bias in the FVTX-MuTr mismatch distribution normalization 
(\ref{eq:norm_mismatch}); or 3) missing accidental hit-track 
associations in simulation. The third Gaussian contribution in 
Eq.~(\ref{eq:detres}), $f_3$, is varied by 20\% to account for these 
uncertainties. This variation produces a change in the \bfrac of up to 
0.004.

\subsubsection{Relative detector acceptance and efficiency.}
\label{sec:acc_eff_sys}

Variations in the {\sc pythia{\footnotesize 8}} parameters, such as 
renormalization factor and additional weighting in order to match 
measured \jpsi rapidity and \pt distributions in \cite{Adare:2011vq}, 
used to determine the relative detector acceptance and efficiency 
$\varepsilon_B/\varepsilon_{J/\psi}$ introduce 1\% relative fluctuation 
in the \bfrac result in the \pp. When considering different scenarios 
for the $B$-meson nuclear modification in Cu+Au; including variations with 
centrality, \pt and rapidity; the fluctuation in \bfrac relative to the 
default result is 5\%.

\subsubsection{Dimuon combinatorial background and FVTX-MuTr mismatch 
distribution normalizations.}
\label{sec:cb_mismatch_norm_sys}

The dimuon combinatorial background normalization (\ref{eq:norm_cb}) 
varies by up to 3\% when changing the mass range used to determine it. 
Another concern is how particle activity surrounding a B~meson can 
affect the normalization (\ref{eq:norm_mismatch}) of FVTX-MuTr mismatch 
distributions. The normalization shows a 5\% variation when embedding 
entire {\sc pythia{\footnotesize 8}}+{\sc geant{\footnotesize 4}} 
events containing prompt \jpsi and \btojpsi. When 
applying these variations in the \dcar fitting (\ref{eq:final_fit}), the 
\bfrac result has a standard deviation of 0.01.

\subsubsection{Correlated background.}
\label{sec:correlated_bg_sys}

The dimuon mass fitting uncertainty for the correlated background 
contributions shown in Table \ref{tab:jpsi_counts} is introduced as 
Gaussian random numbers before each fit. The fraction of \bb in the 
correlated background is varied by f$_{b\bar{b}}=0.32\pm0.21$ in \pp 
collisions based on the uncertainties in the total \cc and \bb cross 
section \cite{Adare:2006hc,Adare:2009ic}. For Cu+Au collisions, 
f$_{b\bar{b}}\in$[0,1] is considered, which accounts for unknown \cc and 
\bb nuclear modifications. When applying these variations in the fitting 
procedure defined in Eq.~(\ref{eq:final_fit}) the \bfrac standard 
deviation is 0.01 in \pp and 0.025 in Cu+Au.

\subsection{Total systematic uncertainties.}
\label{sec:sys_error_total}

\begin{table}[ht]
\caption{\label{tab:sys_errors}Summary of all absolute systematic 
uncertainties on the \bfrac measurement.}
\begin{ruledtabular} 	\begin{tabular}{lcccc}
	source & \multicolumn{2}{c}{\pp} & \multicolumn{2}{c}{Cu+Au} \\
    & {south} & {north} & {south} & {north} \\\hline
    fitting procedure & \multicolumn{2}{c}{$<$0.001} & 0.005 & 0.005 \\
    simulation weighting & 0.003 & 0.002 & 0.005 & 0.005 \\
    \dcar resolution & 0.004 & 0.004 & 0.012 & 0.015 \\
    detector offset & \multicolumn{2}{c}{$<$0.002} & 0.011 & 0.013 \\
    \dcar tail contribution & \multicolumn{2}{c}{$<$0.001} & 0.002 & 0.004\\
    relative acc.$\times$eff. & \multicolumn{2}{c}{$<$0.001} & 0.005 & 0.005\\
    background normalizations & \multicolumn{2}{c}{$<$0.001} & 0.010 & 0.009 \\
	correlated bg & 0.002 & 0.003 & 0.018 & 0.018\\
    fraction of corr. \bb & 0.008 & 0.008 & 0.017 & 0.018 \\
    TOTAL & 0.009 & 0.010 & 0.033 & 0.034 \\
	\end{tabular} \end{ruledtabular} 
\end{table}

Table \ref{tab:sys_errors} summarizes all systematic uncertainty 
contributions. The total systematic uncertainty also listed in the table 
is obtained by varying the parameters as described in the fitting 
function (\ref{eq:final_fit}) simultaneously, assuming all variations 
are independent, and running several independent fits.

\subsection{Other checks.}
\label{sec:other_checks}

Other tests such as variations in the \dcar fitting range, and the use 
of simulated histograms rather than functions as fitting input, provided 
results that are statistically consistent with the default result. No 
additional systematic uncertainties are assigned from these checks.

\section{Results.}
\label{sec:results}

\subsection{Fraction of B~mesons in the \jpsi sample.}
\label{sec:bfrac_result}

\begin{figure}[ht]
	\includegraphics[width=1.0\linewidth]{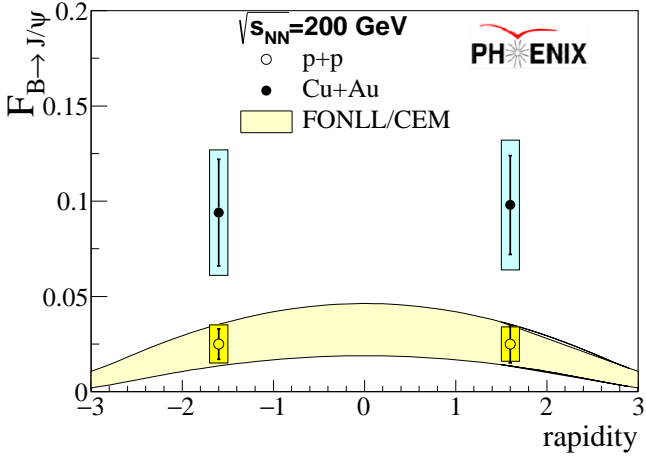}
\caption{\label{fig:bfrac_rap_pp_cuau}Fraction \bfrac of $B$-meson 
decays in the inclusive \jpsi sample in \pp and Cu+Au collisions at 
\full versus rapidity along with a theoretical estimation based on 
fixed-order plus next-to-leading logs (FONLL) 
\cite{Cacciari:1998it,Cacciari:2005rk} for the \btojpsi cross section 
and Color-Evaporation-Model (CEM) \cite{Frawley:2008kk} for the prompt 
\jpsi. The statistical uncertainties are represented by the error bars 
and the systematic uncertainties are represented by filled boxes.}
\end{figure}

\begin{table}[!htb]
\caption{\label{tab:final_results} Fraction of $B$-meson decays in \jpsi 
samples obtained in \pp and Cu+Au collisions at \full.}
\begin{ruledtabular} 	\begin{tabular}{lc}
    data sample & \bfrac \\\hline
    -2.2$<y<$-1.2 \pp & 0.025 $\pm$ 0.008(stat) $\pm$ 0.010(syst)\\
    1.2$<y<$2.2 \pp & 0.025 $\pm$ 0.010(stat) $\pm$ 0.009(syst)\\
    1.2$<|y|<$2.2 \pp & 0.025 $\pm$ 0.006(stat) $\pm$ 0.009(syst) \\ 
\\
    -2.2$<y<$-1.2 (Au-going) & 0.094 $\pm$ 0.028(stat) $\pm$ 0.033(syst)\\
    1.2$<y<$2.2 (Cu+going) & 0.089 $\pm$ 0.026(stat) $\pm$ 0.034(syst) \\
    \end{tabular}  \end{ruledtabular} 
\end{table}

\begin{figure*}[!htb]
\includegraphics[width=0.49\linewidth]{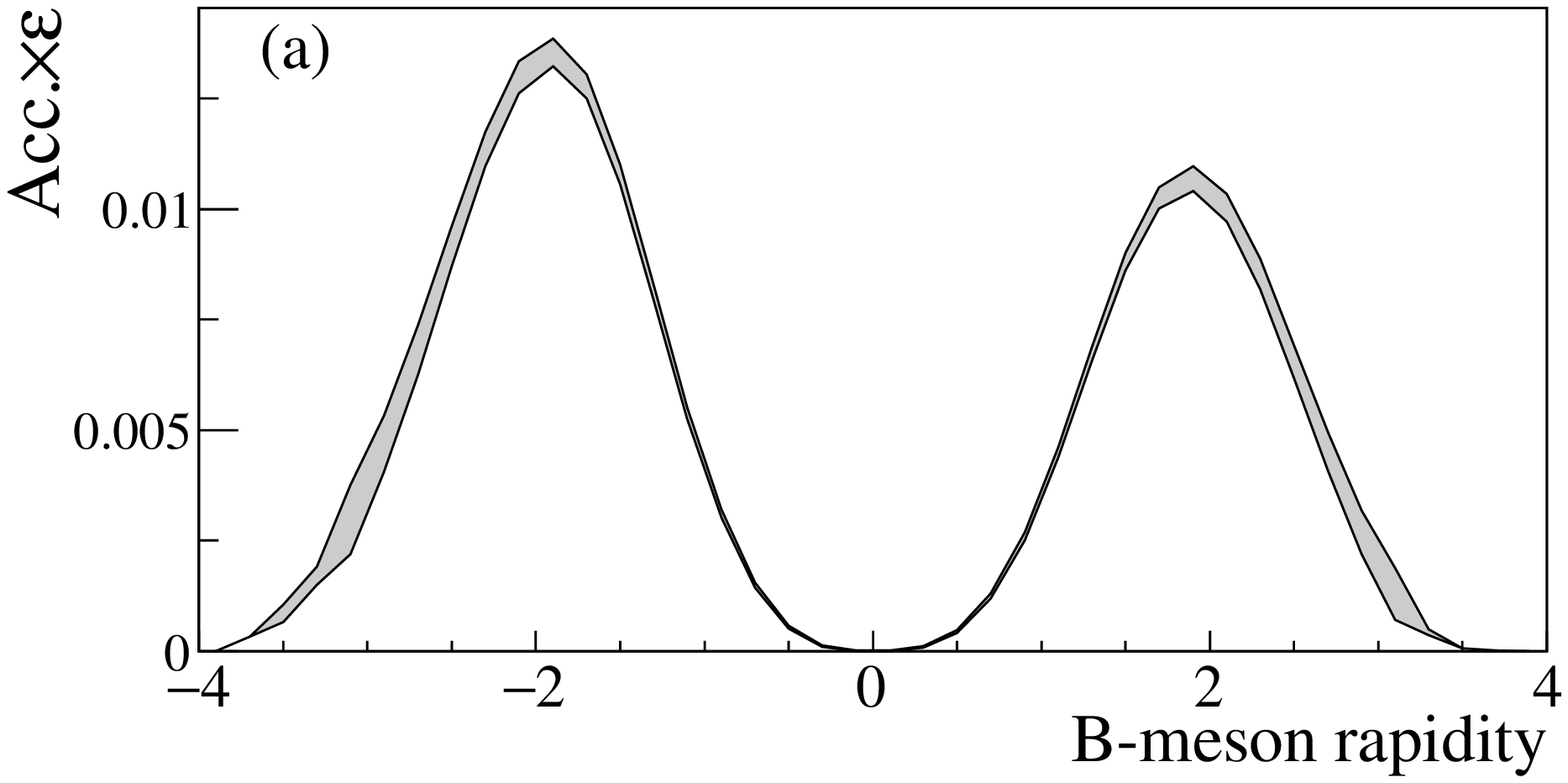}
\includegraphics[width=0.49\linewidth]{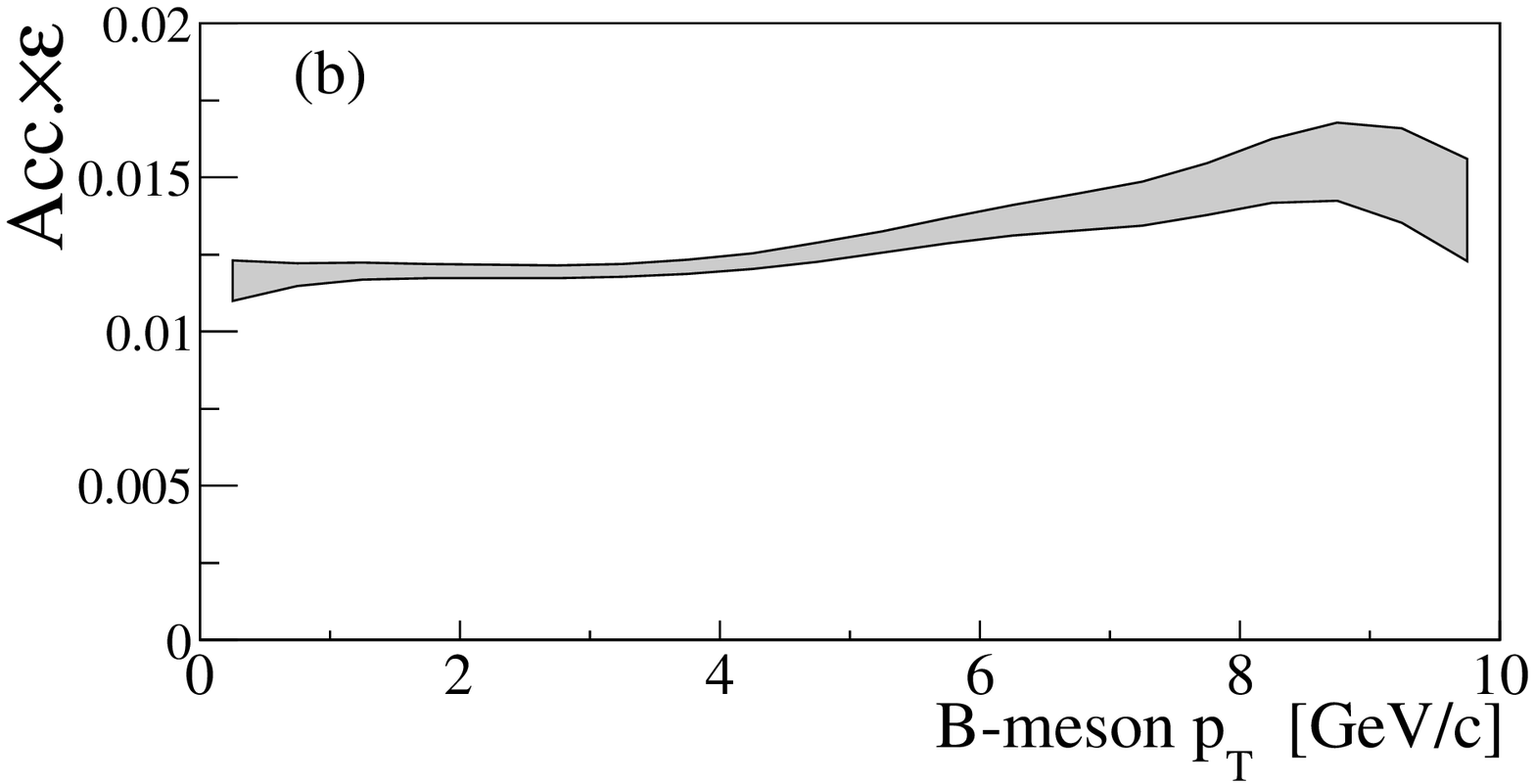}
\caption{\label{fig:bmeson_acceptance} 
Detector acceptance $\times$ efficiency of $B$-meson decays producing a 
\jpsi in the muon arms ($1.2<|y|<2.2$) versus $B$-meson (a) rapidity and 
(b) \pt. The statistical uncertainties in simulation are shown in filled 
gray bands.
}
\end{figure*}

The acceptance and efficiency corrected $B$-meson contributions to the 
\jpsi yields collected in \pp and Cu+Au data are listed in Table 
\ref{tab:final_results} and plotted in Fig. \ref{fig:bfrac_rap_pp_cuau}. 
The detector acceptance and efficiency of $B$-meson decays producing a 
\jpsi in the muon arm apertures is shown in Fig. 
\ref{fig:bmeson_acceptance} as a function of rapidity and transverse 
momentum. The estimation is obtained from the simulation setup described 
in Section \ref{sec:MC_setup} embedded in Cu+Au raw data. As can be seen 
in the right panel, the measurement presented in this analysis is the first
which covers B~mesons starting from zero $p_T$ in heavy ion collisions.

The \bfrac obtained in \pp collisions is well described by a theoretical 
calculation based on 
FONLL~\cite{Cacciari:1998it,Cacciari:2005rk} for the \btojpsi and 
color-evaporation-Model (CEM)~\cite{Frawley:2008kk} for the prompt \jpsi 
differential cross sections.  Uncertainties in the theoretical 
calculation come from the bottom quark mass (4.5 - 5.0 GeV/$c^2$) and 
scale uncertainties. The {\sc cteq6} parton-density 
function~\cite{Nadolsky:2008zw} was adopted in both FONLL and CEM 
calculations.

\subsection{Differential cross section of \bb in \pp collisions.}
The corresponding \pp$\rightarrow$\bb differential cross section can be calculated
by
\begin{equation}
	\label{eq:dsigma_dy}
	\frac{d\sigma_{\bb}}{dy}	= \frac{\frac{1}{2}d\sigma_{J/\psi}/dy \times \bfrac}{\textrm{Br}\left(B\rightarrow J/\psi+X\right)},
\end{equation}

\noindent where $d\sigma_{J/\psi}/dy$ is obtained from \jpsi measurement 
by PHENIX \cite{Adare:2011vq}. The branching ratios 
Br$\left(B\rightarrow J/\psi+X\right)$ and 
Br$\left(J/\psi\rightarrow \mu^+\mu^-\right)$ are reported in the 
particle data group~\cite{Agashe:2014kda}. The factor $\frac{1}{2}$ 
accounts for the fact that both $b$ quarks in the \bb pair contribute to 
the \bfrac. The results are shown in Table \ref{tab:dsigma_dy}, where 
the average rapidity $\left<y\right>$ are obtained from the detector 
acceptance of \btojpsi events generated by {\sc pythia{\footnotesize 
8}}. Table \ref{tab:dsigma_dy} also presents the corresponding FONLL 
calculation.

\begin{table}[htb]
	\caption{\label{tab:dsigma_dy} 
Differential cross section for \pp$\rightarrow$\bb at $\sqrt{s}=$200 GeV 
obtained from (\ref{eq:dsigma_dy}) along with the FONLL theoretical 
calculation~\cite{Cacciari:1998it,Cacciari:2005rk}.}
	\begin{ruledtabular}
    \begin{tabular}{@{\hspace{4em}} cc @{\hspace{4em}} }
    	$\left<y\right>$ & $d\sigma_{\bb}/dy$ [$\mu$b]\\\hline
        -1.6 & 0.51 $\pm$ 0.16(stat) $\pm$ 0.20(sys) \\
        1.6  & 0.52 $\pm$ 0.21(stat) $\pm$ 0.21(sys) \\
        $|1.6|$ & 0.51 $\pm$ 0.13(stat) $\pm$ 0.20(sys)\\
        $|1.6|$ (FONLL) & 0.26$_{-0.10}^{+0.14}$(theory uncert.)\\
    \end{tabular}
    \end{ruledtabular}
\end{table}

\subsection{$B$-meson nuclear modification.}
\label{sec:bmeson_raa}

The nuclear modification factor is defined by
\begin{equation}
	\raa = \frac{\left(dN/{dy}\right)^{\rm CuAu}}
{\left<N_{\rm coll}\right> ~ \left(dN/dy\right)^{pp}},
\end{equation}

\noindent where $dN/dy$ is the yield in Cu+Au and \pp collisions and 
$\left<N_{\rm coll}\right>$ is the average number of binary collisions 
in the Cu+Au data sample. The Glauber estimated average number of 
collisions in Cu+Au collisions is $\left< N_{\rm coll}\right>$=108 $\pm$ 
11. The centrality and $p_T$ integrated \raa for inclusive \jpsi in 
Cu+Au collisions is obtained from the results presented in 
\cite{Aidala:2014bqx}: 0.365 $\pm$ 0.019(stat) $\pm$ 0.026(syst) in the 
Au-going direction and 0.295 $\pm$ 0.026(stat) $\pm$ 0.021(syst) in the 
Cu-going direction with a global uncertainty of 7.1\%. The separated 
prompt \jpsi and $B$-meson \raa can be extracted from inclusive \jpsi \raa 
and \bfrac through
\begin{equation}
\label{eq:bfrac_to_raa}
R_{\rm CuAu}^{\rm prompt} = 
\frac{1-F_{B{\rightarrow}J\psi}^{CuAu}}{1-F_{B{\rightarrow}J\psi}^{pp}} 
R_{\rm CuAu}^{\rm incl.}, ~~
R_{\rm CuAu}^{B} = \frac{F_{B{\rightarrow}/J\psi}^{CuAu}}{F_{B{\rightarrow}J\psi}^{pp}} R_{\rm CuAu}^{\rm incl.},
\end{equation}

\noindent where $R_{\rm CuAu}^{\rm prompt}$, $R_{\rm CuAu}^{B}$ and 
$R_{\rm CuAu}^{\rm incl.}$ are the nuclear modification factors for 
prompt \jpsi, B~mesons and inclusive \jpsi respectively.

\begin{figure}[!ht]
	\includegraphics[width=1.0\linewidth]{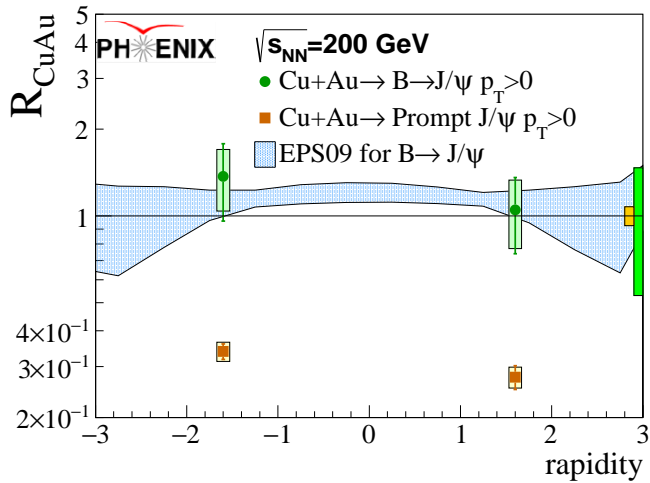}
\caption{\label{fig:bmeson_raa} Rapidity dependence of $B$-meson and 
prompt-\jpsi meson nuclear modification factors \raa along with the 
initial state effect estimated from EPS09 \cite{Eskola:2009uj}.
The statistical uncertainties are shown as bars, and the systematic 
uncertainties are shown as filled boxes.  The boxes at rapidity=3 
are global uncertainties for the prompt \jpsi and \btojpsi.  
}

\end{figure}

The average of the results for \bfrac shown in the first two rows of
Table \ref{tab:final_results} is used as a \pp reference 
$F_{B{\rightarrow}J\psi}^{pp}$. The uncertainties in the \dcar resolution in 
simulation are correlated between the \pp and Cu+Au analysis and cancel 
out in the \raa results. The global uncertainty includes statistical and 
systematic uncertainties of the \pp reference. Table 
\ref{tab:raa_results} and Fig. \ref{fig:bmeson_raa} summarizes the 
$B$-meson and prompt-\jpsi nuclear modifications obtained using 
Eq.~(\ref{eq:bfrac_to_raa}).

\begin{table}[!ht]
\caption{\label{tab:raa_results} Nuclear modification factors of 
B~mesons and prompt \jpsi obtained from Eq.~(\ref{eq:bfrac_to_raa}).}
\begin{ruledtabular} 	\begin{tabular}{lc}
    \multicolumn{2}{c}{$R_{\rm CuAu}^{B}$}\\
    Au-going & 1.37 $\pm$ 0.41(stat) $\pm$ 0.33(syst) $\pm$ 0.47(pp)\\
    Cu+going & 1.05 $\pm$ 0.31(stat) $\pm$ 0.28(syst) $\pm$ 0.47(pp)\\
\\
    \multicolumn{2}{c}{$R_{\rm CuAu}^{\rm prompt}$}\\
    Au-going & 0.339 $\pm$ 0.021(stat) $\pm$ 0.026(syst) $\pm$ 0.075(pp)\\
    Cu+going & 0.276 $\pm$ 0.026(stat) $\pm$ 0.023(syst) $\pm$ 0.075(pp)\\
    \end{tabular} \end{ruledtabular} 
\end{table}

Both nuclear modifications factors from forward and backward rapidities 
are consistent with binary scaling of \pp yields given the large 
uncertainties. The results are also consistent with initial-state 
effects predicted by the EPS09 model \cite{Eskola:2009uj}, shown in Fig. 
\ref{fig:bmeson_raa}, which suggests a modest enhancement. The EPS09 
calculation uses as input $x$ and $Q$ from $gg\rightarrow b\bar{b}$ 
events generated by {\sc pythia{\footnotesize 8}}. The same model 
underpredicts the large yield enhancements observed for leptons from 
inclusive heavy flavor, dominated by charm quarks, at midrapidity and 
negative rapidity in $d$$+$Au 
collisions~\cite{Adare:2013lkk,Adare:2012yxa} at the same energy. Heavy 
flavor yield enhancement at large-$x$, which dominates the negative 
rapidity yield, is also expected from incoherent multiple scattering of 
initial gluons \cite{Kang:2014hha}.

Binary scaling of momentum-integrated heavy flavor production was 
previously observed in electron yields from charm quarks by 
PHENIX~\cite{Adare:2010de} and reconstructed $D$~mesons by 
STAR~\cite{Adamczyk:2014uip} in Au$+$Au collisions at RHIC. The \pt 
integrated $B$-meson nuclear modification result obtained in this work, 
when combined with these earlier results, indicate there is binary 
scaling of $B$- and $D$-mesons separately. Because charm and bottom number 
are conserved in heavy ion collisions at \full, the interaction with the 
QGP medium only alters the momentum distribution of $D$~mesons and 
$B$~mesons causing a relative yield suppression for $p_T\gg m_Q$.

\section{Summary and Conclusions.}

We report the fraction of $B$-meson decays in the inclusive \jpsi yield 
in \pp and Cu+Au collisions at \full. The measurement is centrality and 
\pt integrated with acceptance starting from zero \pt B~mesons. The 
\bb differential cross section obtained from the measured fractions at 
rapidity $1.2<|y|<2.2$ in \pp collisions is consistent with FONLL 
theoretical calculation within 
uncertainties~\cite{Cacciari:1998it,Cacciari:2005rk}. A systematically 
larger fraction is observed in Cu$+$Au collisions than in \pp collisions, 
which reflects a smaller nuclear modification of $B$~mesons in Cu$+$Au 
collisions compared to prompt \jpsi. The nuclear modification factor 
calculated from the Cu$+$Au and \pp fractions, along with the measured 
inclusive \jpsi \raa, are listed in Table \ref{tab:raa_results} and 
shown in Fig. \ref{fig:bmeson_raa}. The results are consistent with 
binary scaling of $B$-meson yields. No significant difference is 
observed between the Cu-going and Au-going direction within the result 
uncertainties.  However, yield enhancement at negative rapidity is 
favored, which is in agreement with cold-nuclear-matter effects observed 
for inclusive heavy flavor in $d$$+$Au collisions at the same 
energy~\cite{Adare:2013lkk,Adare:2012yxa}, an EPS09-based calculation, 
and incoherent multiple scattering of initial gluons.  
This result and others on charm yields indicate that heavy-quark number 
is conserved in heavy ion collisions at \full.  Interaction with the QGP 
medium only alters momentum distributions of $D$ and $B$~mesons. The 
nuclear modification observed for B~mesons contrasts with the strong 
suppression measured for prompt \jpsi~mesons indicating that final-state 
effects, where the \cc binding is broken by the medium formed, are 
dominant for prompt \jpsi~mesons.


\section*{ACKNOWLEDGMENTS}   

We thank the staff of the Collider-Accelerator and Physics
Departments at Brookhaven National Laboratory and the staff of
the other PHENIX participating institutions for their vital
contributions.  We acknowledge support from the
Office of Nuclear Physics in the
Office of Science of the Department of Energy,
the National Science Foundation,
Abilene Christian University Research Council,
Research Foundation of SUNY, and
Dean of the College of Arts and Sciences, Vanderbilt University
(U.S.A),
Ministry of Education, Culture, Sports, Science, and Technology
and the Japan Society for the Promotion of Science (Japan),
Conselho Nacional de Desenvolvimento Cient\'{\i}fico e
Tecnol{\'o}gico and Funda\c c{\~a}o de Amparo {\`a} Pesquisa do
Estado de S{\~a}o Paulo (Brazil),
Natural Science Foundation of China (People's Republic of China),
Croatian Science Foundation and
Ministry of Science and Education (Croatia),
Ministry of Education, Youth and Sports (Czech Republic),
Centre National de la Recherche Scientifique, Commissariat
{\`a} l'{\'E}nergie Atomique, and Institut National de Physique
Nucl{\'e}aire et de Physique des Particules (France),
Bundesministerium f\"ur Bildung und Forschung, Deutscher
Akademischer Austausch Dienst, and Alexander von Humboldt Stiftung (Germany),
National Science Fund, OTKA, EFOP, and the Ch. Simonyi Fund (Hungary),
Department of Atomic Energy and Department of Science and Technology (India),
Israel Science Foundation (Israel),
Basic Science Research Program through NRF of the Ministry of Education (Korea),
Physics Department, Lahore University of Management Sciences (Pakistan),
Ministry of Education and Science, Russian Academy of Sciences,
Federal Agency of Atomic Energy (Russia),
VR and Wallenberg Foundation (Sweden),
the U.S. Civilian Research and Development Foundation for the
Independent States of the Former Soviet Union,
the Hungarian American Enterprise Scholarship Fund,
and the US-Israel Binational Science Foundation.


%
 
\end{document}